\documentclass[11pt]{article}
\usepackage{epsfig}
\usepackage[usenames]{color}
\usepackage{graphicx}
\usepackage{amsmath}
\def\la{\langle}\def\ra{\rangle}
\def\be{\begin{eqnarray}}\def\ee{\end{eqnarray}}
\def\lsim{\mathrel{\rlap{\lower3pt\hbox{\hskip1pt$\sim$}}
     \raise1pt\hbox{$<$}}} 
\def\gsim{\mathrel{\rlap{\lower3pt\hbox{\hskip1pt$\sim$}}
     \raise1pt\hbox{$>$}}} 
\def\le{ \begin{array}{ll}}\def\re{\end{array}}

\def\lear{ \left( \begin{array}{cc}}\def\rear{\end{array} \right)}

\def\le{ \left( \begin{array}{cc}}\def\re{\end{array} \right)}

\def\eft-hls{{\it EFT}$_{\rm bsHLS}$}
\def\skyrmion-hls{{\it Skyrmion}$_{\rm sHLS}$}

\renewcommand{\thefootnote}{\fnsymbol{footnote}}
\begin{document}

\centerline{\Large \bf Scale-Invariant Hidden Local Symmetry,}

\centerline{\Large \bf Topology Change and Dense Baryonic Matter II }
\vskip 0.5cm
\begin{center}

{ Won-Gi Paeng$^a$\footnote{\sf e-mail: wgpaeng@ibs.re.kr},
Thomas T. S. Kuo$^b$\footnote{\sf e-mail: kuo@tonic.physics.sunysb.edu },
Hyun Kyu Lee$^c$\footnote{\sf e-mail: hyunkyu@hanyang.ac.kr},\\
Yong-Liang Ma$^d$\footnote{\sf e-mail: yongliangma@jlu.edu.cn} and
Mannque Rho$^e$\footnote{\sf e-mail: mannque.rho@cea.fr}}

\vskip 0.3cm

{\em $^a$Rare Isotope Science Project, Institute for Basic Science, Daejeon 305-811, Korea}

{\em $^b$Department of Physics and Astronomy, Stony Brook University, Stony Brook, New York 11794, USA }

{\em $^c$Department of Physics, Hanyang University, Seoul 133-791, Korea }

{\em $^d$Center of Theoretical Physics and College of Physics, Jilin University, Changchun, 130012, China }

{\em $^e$Institut de Physique Th\'eorique, CEA Saclay, 91191 Gif-sur-Yvette c\'edex, France }

\vskip 0.3cm
{\ (\today)}
\end{center}

\centerline{ \bf ABSTRACT}

\noindent Exploiting certain robust topological inputs from the skyrmion description  of compressed baryonic matter with a scale-chiral symmetric Lagrangian, we predict the equation of state that is consistent with the properties of nuclear matter at the equilibrium density, supports the maximum mass of massive compact star $\sim 2 M_\odot$ and surprisingly gives the  sound velocity close to the ``conformal velocity"  $1/\sqrt{3}$ at densities $\gsim 3 n_0$.  At the core of this result  is the observation that parity-doubling occurs in the nucleon structure as density goes above $\sim 2n_0$ with a chiral-singlet mass $m_0 \sim (0.6-0.9) m_N$, hinting at a possible up-to-date unsuspected source of proton mass and an emergence at high density of scale symmetry and flavor local symmetry, both hidden in the QCD  vacuum.

\setcounter{footnote}{0}
\renewcommand{\thefootnote}{\arabic{footnote}}
\vskip 1.0cm

\section{Objective}
As to whether certain fundamental symmetries of QCD, invisible or hidden in matter-free space, can emerge at high baryonic density such as in massive compact stars is an extremely interesting but difficult question to address. A variety of emergent symmetries are being discovered in condensed matter systems, and are being discussed in particle physics, including the possibility of emergent gravity and dark matter. In previous papers by the authors~\cite{PKLR,PR-sound}, it was proposed that both hidden local symmetry(HLS) and hidden scale symmetry could emerge at densities relevant to $\sim 2$ solar-mass neutron stars. In this paper, we further sharpen the analysis made in \cite{PKLR}  and explore possible consequences on the phase structure, up to densities hitherto unexplored.

We improve on what was treated in \cite{PKLR} and arrive at a fairly clear resolution of the problem on how the emergent symmetries could manifest in massive compact stars. We should stress that our aim here is basically different from what is being pursued in the astro-nuclear community, which is to obtain with a battery of unknown parameters an equation of state  (EoS) that accommodates the astrophysical observations. Our objective, instead, is to exploit the information provided by astrophysics to zero in on the totally unknown quantity, namely what may be presumed to be highly correlated interactions that enter in dense matter that go beyond what's understood from conventional nuclear systems. What we find is that at the core of the problem is the origin of the proton mass, more than 90\% of which arise out of ``nothing," and the emergence of both the scale and flavor local symmetries that are not visible in the vacuum and baryonic matter at low densities. We admit that our point of view is highly unorthodox in that it deviates, even drastically  at least in appearance, from the currently accepted  paradigm in nuclear physics, namely that anchored on chiral effective field theories involving nucleons and pseudo-Nambu-Goldstone (pNG) bosons.

What we find in a refined analysis of what's  in \cite{PKLR} is that in the same renormalization group (RG) treatment with $V_{lowk}$ adopted in \cite{PKLR}, referred to in what follows as ``$V_{lowk}$-RG,"  the compact star of mass $M_{max}\simeq 2.05 M_\odot$ and radius $R\simeq 12.19$ km with the maximum central density $n_{\rm max}\simeq 5.1n_0$ supports  the sound velocity close to the conformal limit
\be
v_s^2/c^2\simeq 1/3.
\ee
We suggest this to be  a precursor signal for an emergent scale symmetry in dense medium. This result seems to be in strong contrast with what's found in conventional hadronic models belonging to the class of ``energy density functional (EDF),"  $v_c^2/c^2 > 1/3$.

\section{The Effective Lagrangian}
The effective Lagrangian we shall use,   which is  the same as in \cite{PKLR}, referred to as $bs$HLS Lagrangian, consists of the scale-invariant hidden local symmetric term ${\cal L}_{\rm inv}$ plus symmetry breaking term ${\cal L}_{\rm SB}$
\be
{\cal L}={\cal L}_{\rm inv} +{\cal L}_{\rm SB}  \label{TLag}
\ee
where
\begin{eqnarray}
{\cal L}_{\rm inv} &=& {\cal L}_N + {\cal L}_M \, \label{tLag}\\
{\cal L}_N &=&  \bar{N} i \left(\partial_\mu -i g_\rho \vec{\rho}_\mu \cdot \frac{\vec{\tau}}{2} -i g_\omega \frac{\omega_\mu}{2} \right) N - \frac{m_N}{f_\sigma} \chi \bar{N} N
{}+ g_A\bar{N}\gamma^\mu\alpha_{\perp\mu}\gamma_5 N
\nonumber \\
&&
{}+ g_{V\rho}\bar{N}\gamma^{\mu} \left( \alpha_{\parallel\mu} - g_\rho \vec{\rho}_\mu \cdot \frac{\vec{\tau}}{2}\right) N
{}+ g_{V\omega} \bar{N}\gamma^{\mu}\left( \frac{ \partial_\mu \sigma_\omega}{2f_{\sigma\omega}} - g_\omega \frac{\omega_\mu}{2}  \right)N\,,
\label{NLag}    \\
{\cal L}_M &=& \frac{f_\pi^2}{f_\sigma^2} \chi^2 \mbox{tr}\left[ \alpha_{\perp\mu}
  \alpha_{\perp}^{\mu} \right]
{}+ \frac{f_{\sigma\rho}^2}{f_\sigma^2} \chi^2\mbox{tr}\left[ \left( \alpha_{\parallel\mu} - g_\rho \vec{\rho}_\mu \cdot \frac{\vec{\tau}}{2}\right)^2 \right]
{}+ \frac{f_{\sigma\omega}^2}{2f_\sigma^2} \chi^2 \left( \frac{\partial_\mu\sigma_\omega}{f_{\sigma\omega}} - g_\omega \omega_\mu  \right)^2 \nonumber \\
&& {}- \frac{1}{2}\mbox{tr}\left[ \rho_{\mu\nu}\rho^{\mu\nu} \right]
{}- \frac{1}{2}\mbox{tr}\left[ \omega_{\mu\nu}\omega^{\mu\nu} \right]
{}+ \frac{1}{2}\partial_\mu\chi\cdot\partial^\mu\chi  \label{MLag}\\
{\cal L}_{\rm SB} &=& -V(\chi) + \frac{f_{\pi}^2}{4} {\rm Tr} \left({\cal M}U^\dagger +h.c. \right) \left(\frac{\chi}{f_{\sigma}} \right)^3 \label{break}
\end{eqnarray}
where $V(\chi)$ is the scale symmetry breaking potential to be specified below and ${\mathcal M}$ is  the mass matrix which to be consistent with the symmetry we are concerned with, i.e., chiral-scale symmetry, should be of 3 flavors but we will focus on the $SU(2)$ sector, and
\begin{eqnarray}
& \rho^{\mu\nu} = \partial^\mu \vec{\rho}^{\,\nu} \cdot \frac{\vec{\tau}}{2} - \partial^\nu \vec{\rho}^{\,\mu} \cdot \frac{\vec{\tau}}{2}
{}- ig_\rho \left[ \vec{\rho}^{\,\mu} \cdot \frac{\vec{\tau}}{2}, \vec{\rho}^{\,\nu} \cdot \frac{\vec{\tau}}{2} \right]\,, \\
& \omega^{\mu\nu} = \partial^\mu \frac{\omega^\nu}{2} - \partial^\nu \frac{\omega^\mu}{2}
\end{eqnarray} and
\begin{eqnarray}
& \alpha^\mu_\perp = \frac{1}{2i} \left[ \partial^\mu \tilde{ \xi}_R \cdot \tilde{ \xi}_R^{\dagger} - \partial^\mu \tilde{ \xi}_L \cdot \tilde{ \xi}_L^{\dagger}  \right]\,, \\
& \alpha^\mu_\parallel = \frac{1}{2i} \left[ \partial^\mu \tilde{ \xi}_R \cdot \tilde{ \xi}_R^{\dagger} + \partial^\mu \tilde{ \xi}_L \cdot \tilde{ \xi}_L^{\dagger}  \right]
\end{eqnarray} with
\begin{eqnarray}
\tilde{\xi}_{L,R}= e^{i\sigma_\rho/{f_{\sigma\rho}}}e^{\mp i\pi/{f_\pi}}\,.
\end{eqnarray}
The potential $V(\chi)$ is to encode  the trace anomaly of QCD\footnote{We ignore the anomalous dimension of the quark mass operator.},
\begin{equation}
\theta_\mu^\mu=\frac{\beta(\alpha_s)}{4\alpha_s}G^a_{\mu\nu} G^{a\mu\nu} +\sum_{q=u,d,s} m_q\bar{q}q
\end{equation} with the gluon $G$ and the quark $q$. In terms of the effective fields, chiral field $U=\xi_L^\dagger \xi_R$ and  ``conformal compensator" field $\chi$, the trace of the energy-momentum tensor (TEMT) is given by\footnote{We are being cavalier here. In order to get this result, one would have to first make ${\cal L}_{\rm inv}$ scale invariant using the conformal compensator field and accounting for explicit symmetry breaking. This procedure will be explained below with pertinent references.}
\begin{equation}
\theta_\mu^\mu= 4V(\chi) - \chi \frac{\partial V(\chi)}{\partial \chi} + \frac{f_{\pi}^2}{4}{\rm Tr} \left({\cal M}U^\dagger +h.c. \right) \left(\frac{\chi}{f_{\sigma}} \right)^3. \label{TEMT0}
\end{equation}

As in \cite{Paeng:2013xya}, we consider $\rho$ and $\omega$ mesons as the gauge bosons of $\left[ SU(2)_V\times U(1)_V \right]_{local}$ HLS with the gauge couplings $g_{\rho,\omega}$ and $\pi$ meson as the pNG boson of $\left[SU(2)_L\times SU(2)_R \right]_{global}$ chiral symmetry. There is a strong indication in both mean-field~\cite{Paeng:2013xya} and renormalization-group~\cite{PKLR} analyses that global $U(2)$ symmetry for the vector mesons is badly broken at high density. We introduce the Lorentz scalar and iso-scalar field $\sigma$ -- referred to as ``dilaton'' -- as  the Nambu Goldstone boson of the scale symmetry. The fields we are concerned  with transform as
\begin{eqnarray}
\xi_{L,R} &\rightarrow& uh\,\xi_{L,R}\, g_{L,R}^\dagger\,, \\
g_\rho \vec{\rho}_\mu \cdot \frac{\vec{\tau}}{2} &\rightarrow& \lambda h\, g_\rho \vec{\rho}_\mu \cdot \frac{\vec{\tau}}{2}\, h^\dagger - i \partial_\mu h \cdot h^\dagger\,, \\
g_\omega \frac{\omega_\mu}{2} &\rightarrow& \lambda u\, g_\omega \frac{\omega_\mu}{2}\, u^\dagger - i \partial_\mu u \cdot u^\dagger\,, \\
\chi &\rightarrow& \lambda \chi\,, \\
N &\rightarrow& \lambda^{\frac{3}{2}}\, u\,h\, N
\end{eqnarray} under the scale transformation, $x \rightarrow \lambda^{-1} x$, where $g_{L,R} \in \left[SU(2)_{L,R}\right]_{\rm global}$, $h \in \left[ SU(2)_V\right]_{local}$ and $u \in \left[U(1)_V \right]_{local}$. We parameterize $\chi$, the conformal compensator field,  and $\xi_{L,R}$  as
\begin{eqnarray}
\frac{\chi}{f_\sigma} &=& \exp\left(-\frac{\sigma}{f_\sigma}\right)\,,\\
\xi_{L,R} &=& \exp\left(\frac{i\sigma_\omega}{2f_{\sigma\omega}}\right)\, \exp\left(\frac{i\vec{\sigma}_\rho \cdot \vec{\tau}}{2f_{\sigma\rho}} \right)\, \exp\left(\mp\frac{i\vec{\pi}\cdot\vec{\tau}}{2f_{\pi}} \right)\,,
\end{eqnarray} where $\sigma_\omega$ and $\sigma_\rho$ are would-be Nambu Goldstone bosons of HLS that will be Higgsed away and $f_\sigma$, $f_\pi$, $f_{\sigma\omega}$ and $f_{\sigma\rho}$ are the decay constants of the associated fields.

How this simplified Lagrangian is arrived at from a scale-invariant baryonic hidden local symmetric Lagrangian~\cite{LMR-2016} goes as follows. For illustration it suffices to consider (mesonic) nonlinear sigma model as done by Crewther and Tunstall~\cite{CT}.\footnote{A similar procedure with the possible existence of an IR fixed point was proposed by Golterman and Shamir~\cite{GS}. The relation between the two at the leading scale-chiral order is discussed in \cite{MR-omega}.} Including baryon fields is straightforward. Expanded from below, near an IR fixed point $\beta(\alpha_{IR})$, scale-symmetric sigma model Lagrangian to the leading chira-scale order $O(p^2)$ takes the form
\be
{\cal L} &=&{\cal L}_{\rm inv} + {\cal L}_{anom} +V(\chi),\\
{\cal L}_{\rm inv} & = & c_1 {\frac{f_\pi^2}{4} \left(\frac{\chi}{f_\sigma}\right)^2} {\rm Tr} \left(\partial_\mu U\partial^\mu U^\dagger \right)+c_2\frac 12 {f_\sigma^2} \partial_\mu\chi\partial^\mu\chi,\label{inv}\\
{\cal L}_{anom} & = & \Bigg\{(1-c_1) {\frac{f_\pi^2}{4} \left(\frac{\chi}{f_\sigma}\right)^2} {\rm Tr} \left(\partial_\mu U\partial^\mu U^\dagger \right)\nonumber\\
&&\;\;\;{} +(1-c_2)\frac 12 {f_\sigma^2}\partial_\mu\chi\partial^\mu\chi\Bigg\}\left(\frac{\chi}{f_\sigma}\right)^{\beta^\prime},\\
V(\chi) & = & \left(\frac{\chi}{f_\sigma}\right)^4 \left[c_3+c_4\left(\frac{\chi}{f_\sigma}\right)^{\beta^\prime}\right] .
\ee

Following \cite{CT},  we set,  in the chiral limit, $c_1=c_2=1+O(p^2)$ { which can be arrived at by setting the dilaton field equal to zero for processes that do not involve scalar excitations.  The best way to  understand this relation is that there is hidden scale symmetry in Standard Higgs-type Lagrangian that yields both the scale-symmetric form and the nonlinear sigma model form that can be reached when one dials a constant, respectively,  to weak coupling limit and to strong-coupling limit~\cite{yamawaki}.}  Keeping to $O(p^2)$ in the leading-order Lagrangian, we have
\be
{\cal L}_{\rm inv}= {\frac{f_\pi^2}{4} \left(\frac{\chi}{f_\sigma}\right)^2} {\rm Tr} \left(\partial_\mu U\partial^\mu U^\dagger \right)+\frac 12 {f_\sigma^2}\partial_\mu\chi\partial^\mu\chi +O(p^4).
\ee
Hidden-local-symmetrizing this, we have $s$HLS Lagrangian\cite{MR-omega},
\be
{\cal L}_{sHLS} & = & f_\pi^2 \left(\frac{\chi}{f_\sigma}\right)^2{\rm
Tr}\left[\hat{a}_{\perp\mu}\hat{a}_{\perp}^{\mu}\right] + a f_\pi^2  \left(\frac{\chi}{f_\sigma}\right)^2{\rm
Tr}\left[\hat{a}_{\parallel\mu}\hat{a}_{\parallel}^{\mu}\right]\nonumber\\
& &{} - \frac{1}{2g^2}{\rm Tr}\left[V_{\mu\nu}V^{\mu\nu}\right] + \frac{1}{2} \partial_\mu \chi \partial^\mu \chi +O(p^4).\label{shls}
\ee
We have written this Lagrangian in $U(2)$ symmetric way for notational simplicity. It will be broken to $SU(2)\times U(1)$ at high density.

In the presence of the vector mesons, there is an anomalous-parity term called homogeneous Wess-Zumino term (hWZ) which is left out in the above. It does not figure when baryons are present but it needs to be treated in the skyrmion approach and will be commented on at the conclusion section.  The dilaton potential  can be written to the leading order in scale-chiral symmetry as~\cite{LMR-2016}
 \begin{eqnarray}
V(\chi) &\approx   & {} \frac{m_\sigma^2 f_\sigma^2}{4} \left( \frac{\chi}{f_\sigma}\right)^4 \left[\ln \left( \frac{\chi}{f_\sigma}\right) - \frac{1}{4}\right],
\label{eq:potentiallog}
\end{eqnarray}
which is the dilaton potential familiar in the literature valid in the limit $\beta^\prime \ll 1$\footnote{ There is an intriguing indication that $\beta^\prime\sim 2$ in dense nuclear matter~\cite{MR-omega}. This would imply that this approximation may be untenable.}. In the derivation of (\ref{eq:potentiallog}), the mass formula valid in the chiral limit, i.e.,  $m_\sigma^2 f_\sigma^2 = 4\beta^\prime(4 + \beta^\prime)c \simeq 16 \beta^\prime c$ -- which is the dilaton analog to the Gell-Mann-Oakes-Renner relation for the pion -- is used.
Now decomposing $U(2)\to SU(2)\times U(1)$ and coupling baryons in HLS way to the Lagrangian (\ref{shls}) leads to the $bs$HLS Lagrangian (\ref{TLag}).

\section{From Skyrmions to Effective Field Theory}
Our principal thesis is that one can map certain robust properties dependent on topological structure encoded in the skyrmion approach -- with $s$HLS --  to the effective field theory approach  --with $bs$HLS -- in accessing dense baryonic matter. Let us denote the former approach as {\it Skyrmion}$_{\rm sHLS}$ and the latter as {\it EFT}$_{\rm bsHLS}$. Our strategy is to establish a connection between the two, and exploit the power of both approaches to explore the EoS of compact-star matter.

Since the essential ideas are developed in detail elsewhere, we merely summarize them here in a concise way and then focus on sharpening the arguments made previously (\cite{PKLR}) on what and how they will be correlated.

\subsection{Effective field theory with $bs$HLS Laragian}

We assume that the Lagrangian (\ref{TLag}) is defined at a scale $\Lambda_M$,  matched to QCD via current correlators~\cite{HLS,HY:PR}. The matching endows the ``bare" parameters of the effective Lagrangian with nontrivial  dependence on both perturbative and non-perturbative properties of QCD. In particular, it allows the EFT Lagrangian to track the vacuum change in terms of  various condensates, i.e., quark condensate $\Sigma\equiv \la\bar{q}q\ra$, gluon condensate ${\cal G}\equiv\la G^2\ra$ etc. Suppose the Lagrangian is embedded in dense medium. The vacuum change caused by density will then reflect on the change in the condensates involved and hence on the parameters of the EFT Lagrangian. The resulting density dependence, intrinsic of QCD from the QCD-EFT matching,  will be referred to as ``intrinsic density dependence" (IDD for short).  The definition of IDD -- and how it enters in nuclear dynamics -- will depend on how the theory is formulated. This inevitably brings in  certain non-uniqueness in the procedure. In this paper, it will be defined with the $bs$HLS Lagrangian for {\it EFT}$_{\rm bsHLS}$ along the line fully described in \cite{PKLR}. If the massive degrees of freedom are integrated out, as in the standard ChPT (sChPT for short) the IDDs that figure  can effectively contain certain density-dependent higher-order interactions that are integrated out such as short-range $n$-body forces with $n>2$. Thus the IDDs entering into the sChPT could differ from the IDDs of bare $bs$HLS. This point should be kept in mind in following the reasoning developed in this paper.

\subsection{Skyrmions on crystal lattice with $s$HLS Lagrangian}

Given the framework of {\it EFT}$_{\rm bsHLS}$, IDDs can be more or less determined up to nuclear matter density $n_0\simeq 0.16$ fm$^{-3}$ either from experiments or theoretically in sChPT or in the future, perhaps by lattice calculations.  It will however be extremely difficult to track them at higher densities going beyond $n_0$.  Here we rely on certain topological structure present in the skyrmion approach, recently reviewed with extensive references in \cite{Review-china}. Given the daunting mathematical difficulty in fully quantizing skyrmion matter, one can extract only limited information from the skyrmion approach. Fortunately there are  certain robust topological properties  that can be exploited.  In particular, an extremely important  observation in the skyrmion approach is the changeover from skyrmions to half-skyrmions at a density  $n> n_0$. Involving topology, it is robust even though the effect is present in skyrmions put on crystal lattice, which can be justified only at large $N_c$. In fact a hint for the existence of a half-skyrmion structure is already present, although invisible,  in light nuclei such as the $\alpha$ particle. It is found to provide the most important tool to enable one to access compact-star density. The strategy we shall rely on is the skyrmion crystal structure obtained with $s$HLS, i.e.,   ``{\it Skyrmion}$_{\rm sHLS}$.''

When the topology change is translated into the bare parameters of {\it EFT}$_{\rm bsHLS}$, it makes a drastic -- and  novel -- change in the  IDDs across the transition density denoted $n_{1/2}$. Specifically it gives the prediction that going into the half-skyrmion phase, the in-medium pion (dilaton) decay constant $f_\pi^\ast$ ($f_\sigma^\ast$) and the in-medium nucleon mass $m_N^\ast$ go over to a constant
\be
f_\pi^\ast/f_\pi \approx f_\sigma^\ast/f_\sigma\approx m_N^\ast/m_N \to \kappa\label{kappa}
\ee
where $\kappa$ is a (nearly) density-independent constant $\kappa\sim (0.6-0.9)$. We will see that this result is the key ingredient  in making the sound velocity of massive compact stars approach  the ``conformality" $(v_s/c)^2=1/3$. This differs from the predictions given by phenomenological nuclear models, typically  $0.6-0.8$~\cite{haensel}.

The skyrmion crystal prediction (\ref{kappa}) that the effective nucleon mass goes to a constant of order $O(m_N)$ is reminiscent of the parity-doublet nucleon model~\cite{paritydoublet} in which the nucleon mass contains a chiral-invariant mass $m_0$
\be
m_N^\ast = m_0 +\Delta (\Sigma)\label{parity-doubled-mass}
\ee
where $\Delta (\Sigma)\to 0$ as $\Sigma\to 0$. Unlike in the parity-doublet model where $m_0$ is injected {\it ab initio}, however, here it emerges for $n>n_{1/2}$. In the half-skyrmion phase, the quark condensate $\Sigma$ vanishes {\it when space-averaged}, i.e.,  $\overline{\Sigma}\to 0$, but it is non-vanishing locally and hence supporting chiral density waves with a non-vanishing pion decay constant. Therefore chiral symmetry is not really restored as in the case of the parity-doublet structure. The changeover from  skyrmions to half-skyrmions, strictly speaking, is not a bona-fide phase transition with a local order parameter although it behaves very much like one as one also sees in heavy-light hadrons~\cite{ma-harada2016}. Since the pion is present, chiral symmetry is still spontaneously broken.

\subsection{ Mapping {\it Skyrmion}$_{\rm sHLS}$ to {\it EFT}$_{\rm bsHLS}$}\label{mapping}

One of the characteristics of the {\it Skyrmion}$_{\rm sHLS}$ approach to dense matter, i.e., on crystal lattice, is that the solitonic background impacts on -- or ``warps" -- the properties of the degrees of freedom that are involved in the strong interactions. This then makes the parameters  involved -- such as the pion decay constant, the axial-vector coupling constant etc. -- background-dependent.  Now the background depends on density, so those parameters will inevitably slide with density. One may be able to formulate this phenomenon in terms of what is known as ``Klein-Kaluza metric (warping) effect" due to the background as suggested in \cite{harland}. At very low density and energy, the background effect will be small and hence the Skyrme model -- with pions only -- (consider it simplified from $s$HLS) should be equivalent to chiral perturbation theory, say, sChPT -- with nucleons and pions only (which can be considered as simplified from $bs$HLS). However as density goes up high, the metric warping, particularly with the massive degrees of freedom, could become highly nonlinear, and hence could not be captured by higher order calculations in sChPT.  We are proposing that this warping is mostly, if not wholly, captured in the IDDs in $bs$HLS.

Making a realistic connection between the two -- which we claim must exist -- would require working out quantum theories. Accessing the skyrmion matter quantum mechanically is still, however, far from feasible whereas given the IDDs across $n_{1/2}$, accessing the latter is feasible. We will therefore extract from the skyrmion approach the properties based on topology discussed above that we deem robust and incorporate them in the {\it EFT}$_{\rm bsHLS}$ approach.  More specifically the relations (\ref{kappa}) and (\ref{parity-doubled-mass})  will be imported into doing the first decimation of the $V_{lowk}$ RG.

\section{Analysis in the Mean-Field Approximation}\label{mean field}\label{MFT}
\label{meanfield}
We first apply the $bs$HLS Lagrangian (\ref{TLag}) to dense matter in the mean-field approximation. Relativistic mean-field (RMF) approach has been extensively used in nuclear physics for  both finite and infinite systems. Its overall success for finite and infinite systems up to nuclear matter density can be understood as an indication that doing RMF calculation is equivalent to doing Landau Ferrmi-liquid fixed point theory~\cite{matsui}. Stated in terms of effective field theory with Fermi surface,  the approximation would become more accurate as density increases. This is because in Wilsonian RG approach,  the Fermi-liquid fixed point is approached as $\tilde{\Lambda}/k_F\to 0$ where {$\tilde{\Lambda}=\Lambda-k_F$} ($\Lambda$ being the cutoff scale on top of the Fermi surface from which the decimation is done), provided of course, as we assume, the Fermi-liquid structure continues to hold.

The thermodynamic potential $\Omega = E -TS - \mu N$ divided by the volume $V$ at zero temperature  in mean-field with (\ref{TLag}) for symmetric nuclear matter is
\begin{eqnarray}
\left. \frac{\Omega(T=0)}{V} \right|_{\omega_0 = \langle \omega_0 \rangle,\, \chi = \langle \chi \rangle}
&=& \frac{1}{4\pi^2} \left[ 2E_F^3 k_F - m_N^{\ast\,2}E_F k_F - m_N^{\ast\,4} \ln\left( \frac{E_F + k_F}{m_N^\ast}\right) \right] +  V(\langle \chi \rangle) \nonumber\\
&& + \left[ g_\omega \left( g_{V\omega} -1 \right) \langle \omega_0 \rangle -\mu \right] \frac{2}{3\pi^2} k_F^3 -\frac{1}{2} f_{\sigma \omega}^2 g_\omega^2 \frac{\langle \chi \rangle^2}{f_\sigma^2} \langle \omega_0 \rangle^2\label{Omega}
\end{eqnarray} where $\langle \omega_0 \rangle$ and $\langle \chi \rangle$ are the vacuum expectation value(VEV)of $\omega_{\mu =0}$ and $\chi$, $m_N^\ast = \frac{\la\chi\ra}{f_\sigma} m_N $ and $E_F = \sqrt{k_F^2 + m_N^{\ast\,2}}$. The nucleon number density is
\begin{equation}
n \equiv N/V =  -\frac{\partial (\Omega/V)}{\partial \mu} = \frac{2}{3\pi^2} k_F^3
\end{equation}
and the chemical potential $\mu$ given by the condition  $\frac{\partial (\Omega/V)}{\partial n} = 0$
is
\begin{equation}
\mu = E_F + g_\omega \left( g_{V\omega} -1 \right) \langle \omega_0 \rangle\,.
\end{equation}
The energy density $\epsilon$ and the pressure $P$ at $T=0$ are given by
\begin{eqnarray}
\epsilon
&=& \frac{1}{4\pi^2} \left[ 2E_F^3 k_F - m_N^{\ast\,2}E_F k_F - m_N^{\ast\,4} \ln\left( \frac{E_F + k_F}{m_N^\ast}\right) \right]  \nonumber\\
&& + g_\omega \left( g_{V\omega} -1 \right) \langle \omega_0 \rangle n -\frac{1}{2} f_{\sigma \omega}^2 g_\omega^2 \frac{\langle \chi \rangle^2}{f_\sigma^2} \langle \omega_0 \rangle^2 +  V(\langle \chi \rangle) \label{eden}
\end{eqnarray}
and
\begin{eqnarray}
P &=& -\left. \frac{\Omega}{V} \right|_{\omega_0 = \langle \omega_0 \rangle,\, \chi = \langle \chi \rangle} \\
&=& \frac{1}{4\pi^2} \left[ \frac{2}{3}E_F k_F^3 - m_N^{\ast\,2}E_F k_F + m_N^{\ast\,4} \ln\left( \frac{E_F + k_F}{m_N^\ast}\right) \right] \nonumber\\
&& + \frac{1}{2} f_{\sigma \omega}^2 g_\omega^2 \frac{\langle \chi \rangle^2}{f_\sigma^2} \langle \omega_0 \rangle^2 -  V(\langle \chi \rangle)\,.\label{pre}
\end{eqnarray}

Before proceeding, a side remark is in order here regarding the thermodynamic consistency and IDDs.

Given the IDD in (\ref{TLag}) embedded in medium, the parameters $g_{V\omega}$, $g_\omega$,  $f_{\sigma \omega}$ and the hadron masses are density dependent. In obtaining the above thermodynamics relations, we have cavalierly ignored dependence on density in those parameters in taking partial derivatives with respect to density.  For thermodynamic consistency, however, it is necessary to take into account the density dependence in terms of local baryon field operators as was discussed in \cite{chaejun}.  A naive manipulation of density as a c-number in writing down the equations of motion would bring inconsistency to thermodynamics relations. This has to do with what's known in nuclear theory as ``rearrangement terms." They are associated in nuclear theory with many-body correlations. What this implies is that there is an ambiguity in defining IDDs. How the IDDs are defined depends on at what scale the ``matching" between EFT and QCD is done and what the relevant degrees of freedom in EFT are at that matching scale and how higher-order (e.g., loop) corrections are incorporated. An example is the effect of $n$--body forces for $n > 2$ in EFT.

Suppose that the RG decimation is done from the scale $\Lambda$ picked below the $\omega$ mass, say $\sim 400-500$ MeV  as in standard chiral perturbation (sChPT) approaches (see e.g. \cite{holt-kaiser}). Then the IDDs figuring in sChPT Lagrangian defined at $\Lambda_{EFT}$ with the $n$-body potentials for $n >2$ ``integrated out" will contain not only the effects inherited from QCD at the matching scale $\Lambda_M > \Lambda$ but also those effects decimated out from $\Lambda_M$ to $\Lambda$. The latter will involve effects of $n$-body forces that involve mass scales of vector mesons. For instance, short-range three-body-force effects could enter into the {\it effective} IDDs of sChPT Lagrangians.  This is one way to understand how the famous C14 dating process can be explained more or less equally well  (i) with IDDs but  without three-body forces and (ii) with three-body forces and without IDDs~\cite{holt-weise-c14}.  In \cite{chaejun}, this problem was resolved by treating the density as the VEV of the bilinear nucleon field operator $N^\dagger N$, and the field operator explicitly taken into account in writing down equations of motion. This reflects on that what's involved is higher-dimension effective field operators in the Lagrangian that in the mean field are given in terms of the VEV of density operator.
The upshot of this somewhat intricate and fuzzy relation is, however, that when done correctly, it should turn out that at the Fermi-liquid fixed point, the parameters should depend on the Fermi-momentum.

Returning to the main flow of the discussion, we look at the stationarity  conditions that give the gap equations for $\chi$ and $\omega$
\begin{eqnarray}
\left. \frac{\partial\, \Omega }{\partial \chi} \right|_{\omega_0 = \langle \omega_0 \rangle,\, \chi = \langle \chi \rangle} = 0\,, \quad \left. \frac{\partial \,\Omega }{\partial \omega_0} \right|_{\omega_0 = \langle \omega_0 \rangle,\, \chi = \langle \chi \rangle} = 0\,.
\end{eqnarray} They lead to
\begin{eqnarray}
&\frac{m_N^2\langle \chi \rangle}{\pi^2 f_\sigma^2} \left[ k_F E_F - m_N^{\ast\,2} \ln \left( \frac{k_F + E_F}{m_N^\ast} \right) \right] -\frac{f_{\sigma\omega}^2}{f_\sigma^2} g_\omega^2 \langle \omega_0\rangle^2 \langle \chi \rangle + \left. \frac{\partial\, V(\chi)}{\partial \chi} \right|_{\chi = \langle \chi \rangle} =0 \,, \label{gap1}\\
&g_\omega \left(g_{V\omega}-1 \right)n -f_{\sigma \omega}^2 g_\omega^2 \frac{\langle \chi \rangle^2}{f_\sigma^2} \langle \omega_0 \rangle = 0\,. \label{gap2}
\end{eqnarray}
One obtains\footnote{Unless otherwise stated, we will work in the chiral limit.} from (\ref{gap1}) and (\ref{gap2}) the VEV of the trace of energy-momentum tensor $\theta_\mu^\mu$
\begin{eqnarray}
\langle \theta^\mu_\mu \rangle
&=& \langle \theta^{00} \rangle - \sum_i \langle \theta^{ii}\ra = \epsilon - 3 P\nonumber\\
& =& 4V(\langle \chi \rangle) - \langle \chi \rangle \left. \frac{\partial V( \chi)}{\partial \chi} \right|_{\chi = \langle \chi \rangle}. \label{TEMT1}
\ee
This is just what one gets by taking the mean-field value of (\ref{TEMT0}). This shows that the Fermi surface does not spoil scale symmetry. In fact we will arrive at the conclusion that strongly-correlated hadronic interactions do not modify the dilaton potential. This feature will account for the emergent scale symmetry in compact-star matter.

\section{Renormalization Group Treatment with $V_{lowk}$}

The treatment given in the mean field in Section \ref{mean field}  corresponds, albeit approximately but surprisingly efficiently, to Landau-Fermi liquid fixed point theory. It would be more reliable --  provided there is no phase change -- as density increases beyond $n_0$. To go beyond the Fermi-liquid approximation, the renormailzation-group appraoch with $V_{lowk}$  ($V_{lowk}$-RG) proves to be most powerful with the scheme developed as in \cite{dongetal,PKLR}. Arriving at the Fermi-liquid fixed point corresponds  to doing what is identified as the ``first decimation" in ~\cite{BR:DD} in terms of  the $V_{lowk}$-RG ~\cite{vlowk-fermiliquid}. Going beyond the fixed point structure, i.e., the ``second decimation," involves sophisticated high order correlation calculations. In what follows, we will simply follow the procedure used in \cite{dongetal,PKLR} to do the two-decimation calculation. There will be nothing new in formalism here. What's new is in the way the IDDs are implemented in the effective ($bs$HLS) Lagrangian (\ref{TLag}).

{\it Our basic premise is that when embedded in the medium characterized by density $n$, formally interpreted in terms of baryon field operators as described above, the Lagrangian preserves the same symmetry structure --  apart from the non-Lorentz covariance -- as in the medium-free vacuum and as stated in Section \ref{mapping},  the effect of changes in the vacuum structure caused by density, including the topology change, is entirely encoded in the way the parameters of the Lagrangian behave as density changes.  Needless to say, should there be phase transitions along the way, this scheme will break down. We will assume that this does not happen in the density domain relevant in compact stars.}

How the IDDs are defined in terms of the properties of scale symmetry involving the dilaton $\chi$ and chiral symmetry involving $\pi$ and $V$ is spelled out in detail in \cite{PKLR}. We will follow what's done there. The first thing to do is then to incorporate the IDD structure  of the ``bare parameters" of the Lagrangian\footnote{We recall that since the RG (first) decimation is done from $\Lambda < \Lambda_M$, certain many-body induced effects will go into the ``effective" IDDs. This will make the effective IDDs most likely different from the IDDs given by the matching at $\Lambda_M$.}.  The Lagrangian that gives both nucleon and meson masses in  (\ref{TLag}) is
\begin{equation}
{\cal L}_{\rm mass} = - \frac{m_N}{f_\chi} \chi \bar{N} N + \frac{f_{\sigma\rho}^2}{2f_\chi^2} \chi^2 \left( g_\rho \rho_\mu^a \right)^2
+ \frac{f_{\sigma\omega}^2}{2f_\chi^2} \chi^2 \left( g_\omega \omega_\mu  \right)^2 + {\cal L}_{\rm SB} \label{massL}
\end{equation} with the iso-spin index $a$. When embedded in medium, after shifting $\sigma \rightarrow \langle \sigma \rangle + \sigma$ with $\langle \chi \rangle \equiv f_\sigma \exp\left( -\frac{\langle \sigma \rangle}{f_\sigma}\right)$, the masses are given -- with the $\ast$ denoting density dependence -- as
\begin{equation}
{\cal L}_{\rm mass} = -m_N^\ast \bar{N} N + \frac{m_\rho^{\ast\,2}}{2} \left( \rho_\mu^a \right)^2
+ \frac{m_\omega^{\ast\,2}}{2} \left( \omega_\mu  \right)^2 - \frac{m_\sigma^{\ast\,2}}{2} \sigma^2 - \frac{m_\pi^{\ast\,2}}{2} \pi^{a\,2}\,,
\end{equation}
where $\sigma$ and $\pi$ are redefined as $\sigma \equiv \frac{\langle \chi \rangle}{f_\sigma} \sigma  $ and $\pi^a \equiv \frac{\langle \chi \rangle}{f_\sigma} \pi^a $ to get the kinetic term in the form as
\begin{equation}
{\cal L}_{\rm scalar} = \frac{1}{2} \partial_\mu \sigma \partial^\mu \sigma + \frac{1}{2} \partial_\mu \pi^a \partial^\mu \pi^a\,.
\end{equation} Then, the masses are related to $\langle \chi \rangle$, $g_\rho$ and $g_\omega$ -- all of which are carrying IDDs -- as
\begin{equation}
\frac{m_N^\ast}{m_N} \approx \frac{g_V}{g_V^\ast} \frac{m_V^\ast}{m_V} \approx \frac{m_\sigma^\ast}{m_\sigma} \approx \left( \frac{m_\pi^\ast}{m_\pi} \right)^2 \approx \frac{f_\pi^\ast}{f_\pi} \approx \frac{\langle \chi \rangle^\ast}{f_\sigma}\equiv \Phi^\ast\,,\label{masterIDD}
\end{equation} where
\begin{equation}
m_\sigma^{\ast\,2} \equiv - \left. \frac{\partial^2 V(\sigma)}{\partial\,\sigma^2} \right|_{\sigma = \langle \sigma \rangle^\ast}\,.
\end{equation}
This result, first obtained in \cite{PKLR}, is the principal element in our approach with the crucial input on  how the various parameters related in the specific way given in (\ref{masterIDD}) flow as density increases in compact-star matter. In (\ref{masterIDD}),  we used `$\approx$' to indicate (inevitable) small differences between various different degrees of freedom depending on the choice of the cutoff from which the decimations for $V_{low\,k}$ are made. Specifically, the cutoff values for the nucleon and mesons are  taken differently  to account for differences that come from higher-order correlations as mentioned above for,  e.g., three-body forces.

\subsection{The intrinsic density dependence (IDD)}
\begin{figure}[ht]
\begin{center}
\includegraphics[width=7.5cm]{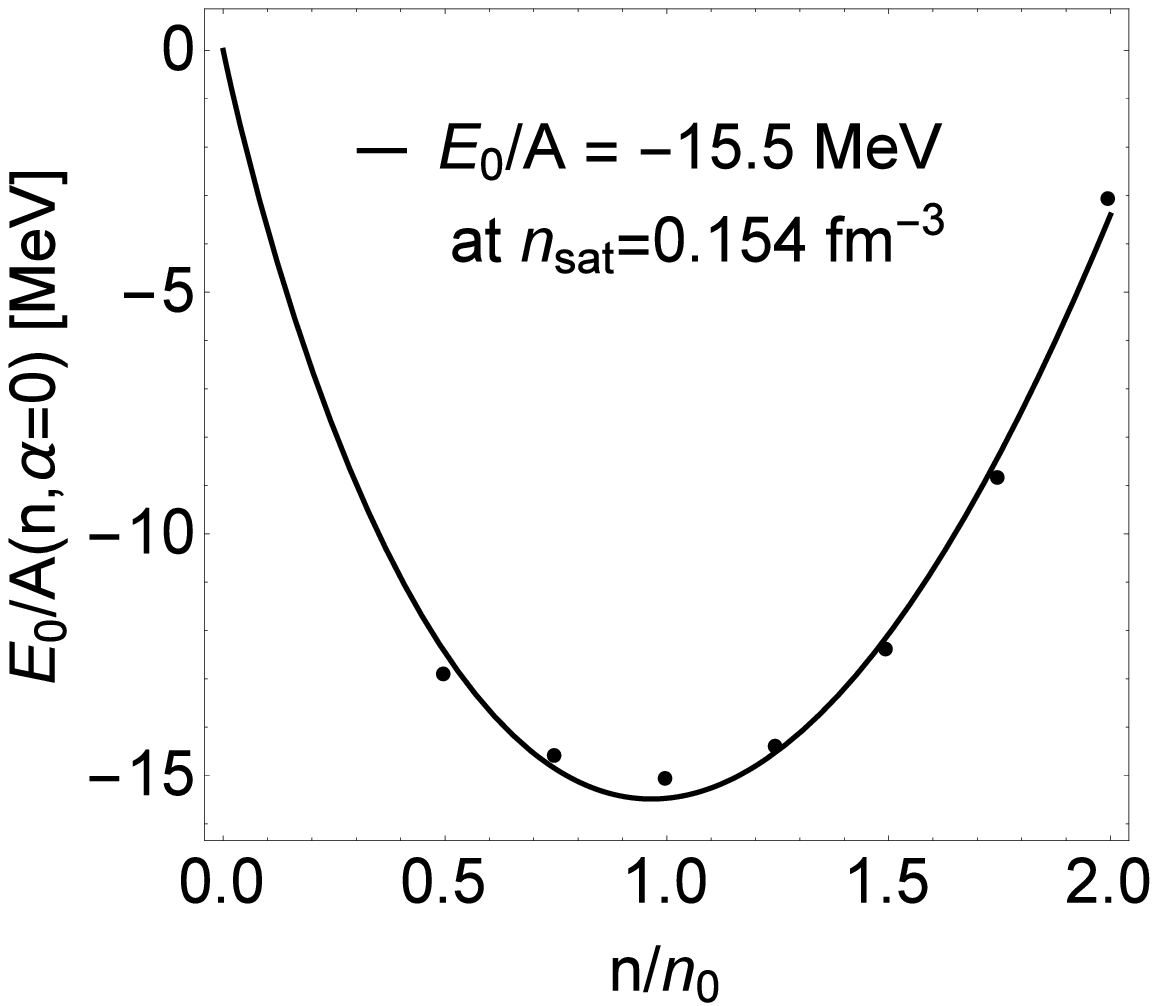}\includegraphics[width=7.5cm]{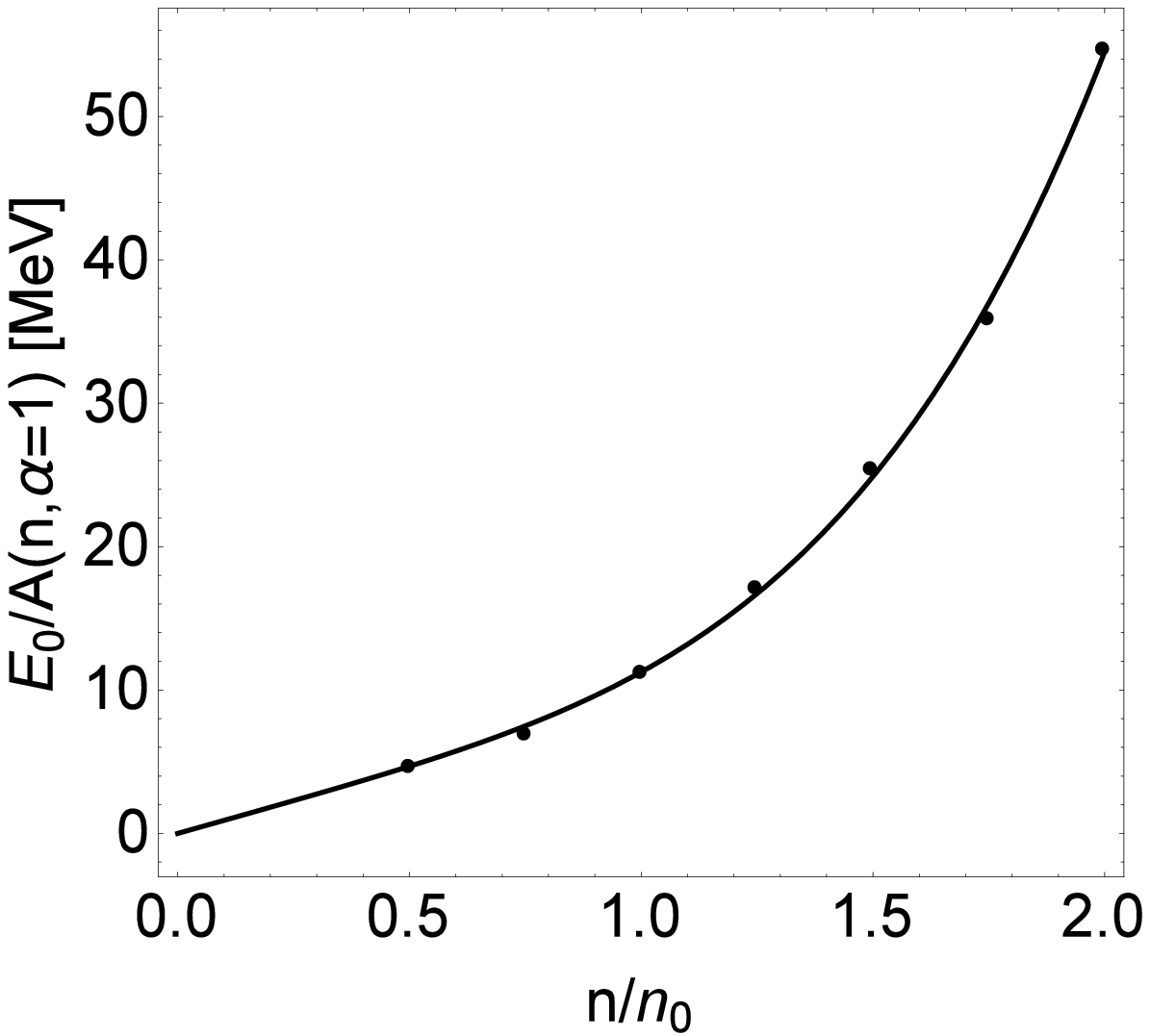}
\caption{ $E_0/A$ for $\alpha =0$(left) and $\alpha = 1$(right) matter, where $\alpha = (N_n - N_p)/A$ and $A=N_n + N_p$ with the proton/neutron number $N_{p/n}$. The dots are from the $V_{lowk}$ RG and the solid lines are fit by Eq.~(\ref{fit}). The saturation density is estimated as $n_{\rm sat} = 0.154 {\rm fm}^{-3}$.
 }\label{groundE0}
 \end{center}
\end{figure}
The only element of the present treatment that differs from  what's done in \cite{PKLR} is the refinement of the behavior of the IDD in the $V_{lowk}$ RG treatment.  Based on the skyrmion matter treatment as discussed above,  we have  an important change in the density dependence in IDD at a density near $n_{1/2} \sim 2n_0$ at which the skyrmion-half-skyrmion changeover takes place. It is this changeover that will be fine-tuned in this analysis. For convenience, we recall the properties of IDDs from \cite{PKLR}:
\begin{enumerate}

\item For $n\leq n_{1/2}$, the only scaling parameter is $\Phi^\ast$. Its precise density dependence is unknown. However there is information available from sChPT calculations backed by date from  deeply bound pionic nuclear systems available up to $\sim n_0$. In the absence of precise analytical form, we adhere to simplicity and parameterize it as
\be
\Phi^\ast = \frac{1}{1+ c_I\, \left( n/n_0 \right)} \ \ {\rm for} \ \ n\leq n_{1/2} \label{chiI}
\ee
and extrapolate it from $n_0$ up to $n_{1/2}$ with a $c_I$ determined at $n_0$.
The coupling constants $g_{\phi NN}$ for $\phi=\sigma,\rho, \omega$ do not scale as argued in \cite{PKLR}. The constant $c_I$ will be fine-tuned between 0.13 and 0.20 as explained in \cite{PKLR} so as to give the ground-state properties of nuclear matter.\footnote{$c_I\simeq 0.20$ at $n_0$ reproduces the empirical value for the pion decay constant $f_\pi^\ast/f_\pi\simeq 0.8$ extracted from deeply bound pionic nuclei. We should stress that within the highly limited number of the free parameters, one can obtain a remarkable description of nuclear matter in this formalism. This is of course not the objective of this paper.} It turns out that the results of the $V_{lowk}$ calculation for $n\lsim n_{1/2}$ can be extremely well fit for both the symmetric nuclear matter and neutron matter by the fitting functions
\begin{equation}
E_0/A = A_I\left( \frac{n}{n_0}\right) + B_I \left( \frac{n}{n_0}\right)^{D_I}.\label{fit}
\end{equation}
They are shown in Fig.~\ref{groundE0}.\footnote{These are obtained -- with high precision -- with the constants $(A_I, B_I, D_I) = (-45.5\ {\rm MeV}, 30.1\ {\rm MeV}, 1.54)$ for $\alpha=0$ (symmetric nuclear matter) and $(A_I, B_I, D_I) = (9.11\ {\rm MeV}, 2.14\ {\rm MeV}, 4.08)$ for $\alpha=1$ (pure neutron matter).}
For completeness, we give the predictions for the symmetric nuclear matter at $n_0$: Equilibrium density $n_0 = 0.154$ fm$^{-3}$,  binding energy BE $=15.5$ MeV, compression modulus $K= 215.2$ MeV.
\begin{figure}[h]
\begin{center}
\includegraphics[width=8.0cm]{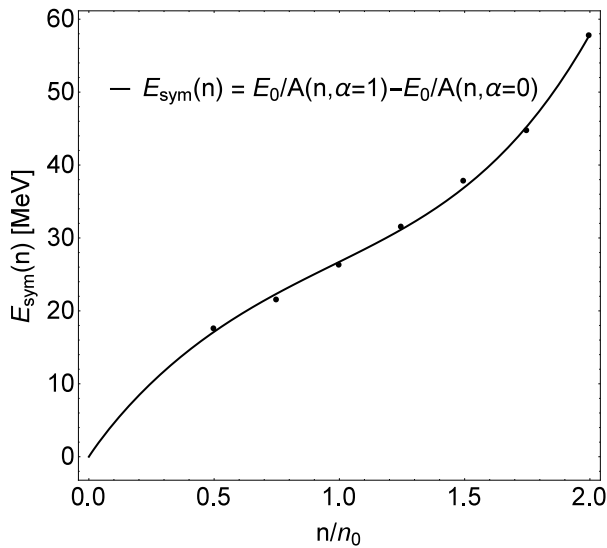}
\caption{ The symmetry energy vs. the density
}\label{symRI}
\end{center}
\end{figure}

The symmetry energy, an extremely important quantity for compact stars, defined by
\begin{equation}
E_{sym} = E_0(n,\,\alpha=1)/A - E_0(n,\,\alpha=0)/A\,,
\end{equation}
can also be fairly accurately calculated. The result is given in Fig.~\ref{symRI} .
Often cited in the literature as ``constraints" for the EoS for compact stars are $E_{sym}|_{n=n_0}$, $L=3n \frac{d E_{sym}}{dn}|_{n=n_0}$, and $K_{sym}= 9 n^2 \frac{d^2 E_{sym}}{dn^2}|_{n=n_0}$:  $E_{\rm sym}/{\rm MeV} = 32 \pm 2$ and $L/{\rm MeV} = 50 \pm 15$.\footnote{ What's given here is a rough set of data. In our view, these ``constraints" given at $n=n_0$ are not  the necessity for getting the EoS at densities relevant to massive compact stars. This point is made in the text.} The predicted values are $E_{sym}\approx 26$ MeV and $L\approx 49$ MeV. The former is a bit smaller than,  while the latter is consistent with, what's quoted in the literature. The $K_{sym}$ cannot be calculated reliably because being a double derivative it is extremely sensitive to the fitting function. In  fact in the literature, various different models give $K_{sym}$ between  - 136 MeV and + 73 MeV~\cite{haensel}. We consider it futile to attempt to pin it down with our scheme that focuses on high-density matter.
\item In the density regime $n > n_{1/2}$, there is a topology change in {\it Skyrmion}$_{\rm sHLS}$ that drastically affects the IDDs at $n=n_{1/2}$.  The existence of such a topology change is one of the most robust inputs from skyrmion matter. From (\ref{kappa}),  we have
\begin{equation}
\Phi^\ast  \approx  \kappa \approx \frac{1}{1+ c_I\, \left( n_{1/2}/n_0 \right)}\ \ {\rm for} \ \ n > n_{1/2}. \label{chiII}
\end{equation}  It is a density-independent constant related to the chiral invariant mass $m_0$ of the nucleon and more significantly to the dilaton condensate $\la\chi\ra$.

Among other IDDs, the most crucial in our approach  is the hidden gauge coupling $g_\rho$ that governs the IDD of the $\rho$ mass by the low-energy theorem $m_\rho\propto f_\pi g_\rho$ with $f_\pi\approx f_\sigma$ in medium. The skyrmion description of the cusp form of the symmetry energy at $n=n_{1/2}$, which is reproduced by the change in the nuclear tensor forces~\cite{cusp}, combined with the vector manifestation(VM) fixed-point structure of HLS leads to that for $n > n_{1/2}$ the coupling $g_\rho$ should drop to zero toward the putative VM fixed point $n_{\rm VM}$. In \cite{dongetal,PKLR}, this feature was approximately represented by the linear form  $g_\rho^\ast/g_\rho \approx 1-n/n_{\rm VM}$.  Below we will modify this scaling in such a way to reproduce the behavior of  the trace of the energy-momentum tensor (TEMT) tending toward a density-independent constant for $n > n_{1/2}$.

As for other parameters, we take them to be as given in \cite{PKLR} for $n > n_{1/2}$,
 \begin{equation}
 \left. \frac{m_i^\ast}{m_i} \right|_{i= N,\, \sigma,\, \rho,\, \omega} = \left(0.71,\, 0.75,\, \frac{g_{\rho NN}^\ast}{g_{\rho NN}},\, 0.73\, \sqrt{\frac{a_\omega^\ast}{a_\omega}} \frac{g_{\omega NN}^\ast}{g_{\omega NN}} \right)\,,
 \end{equation}
 where
 \be \frac{a_\omega^\ast}{a_\omega} &=& \frac{1}{2} \left(1+ \frac{1}{1+0.011(n-n_{1/2})/n_0} \right),\nonumber\\
 \frac{g_{\omega NN}^\ast}{g_{\omega NN}} &=& \frac{1}{1+0.075(n-n_{1/2})/n_0}
 \ee
 which are slightly different from the scalings of $ \frac{a_\omega^\ast}{a_\omega}$ and $\frac{g_{\omega NN}^\ast}{g_{\omega NN}}$ in \cite{PKLR} and compared with those in \cite{PKLR} in Fig. \ref{mVgV}.

 Notable in the analysis there as well as in this paper was that the $U(2)$ symmetry for $\rho$ and $\omega$ which holds fairly well in the vacuum -- and most likely in low-density regime below $n_{1/2}$ -- must break down at higher density. This breakdown requires some small minor modifications from what's taken in \cite{PKLR}. We will explain how this $U(2)$ breakdown can be understood with an emerging scale symmetry in dense medium.

\end{enumerate}
\begin{figure}[h]
\begin{center}
\includegraphics[width=7.5cm]{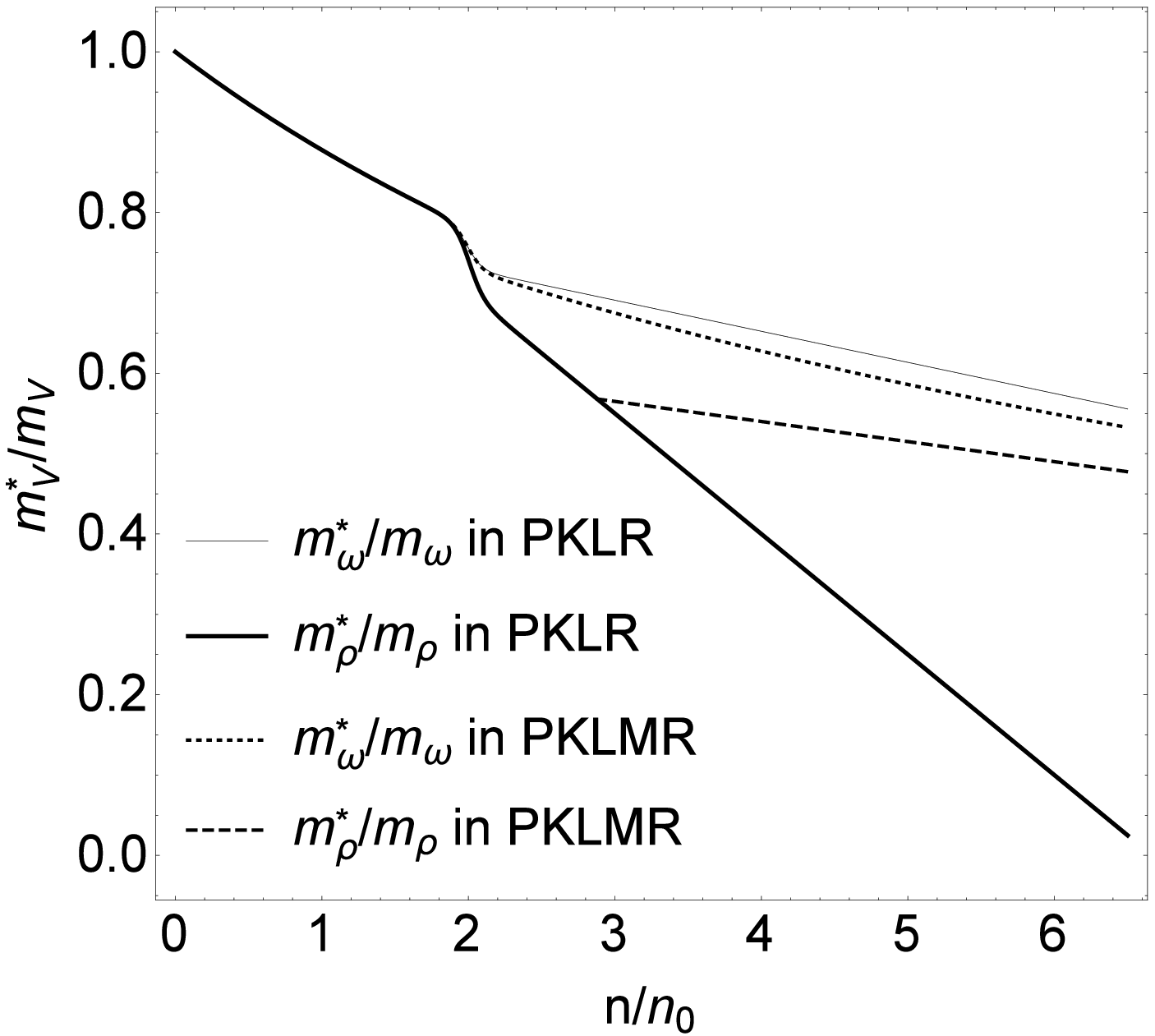}\includegraphics[width=7.5cm]{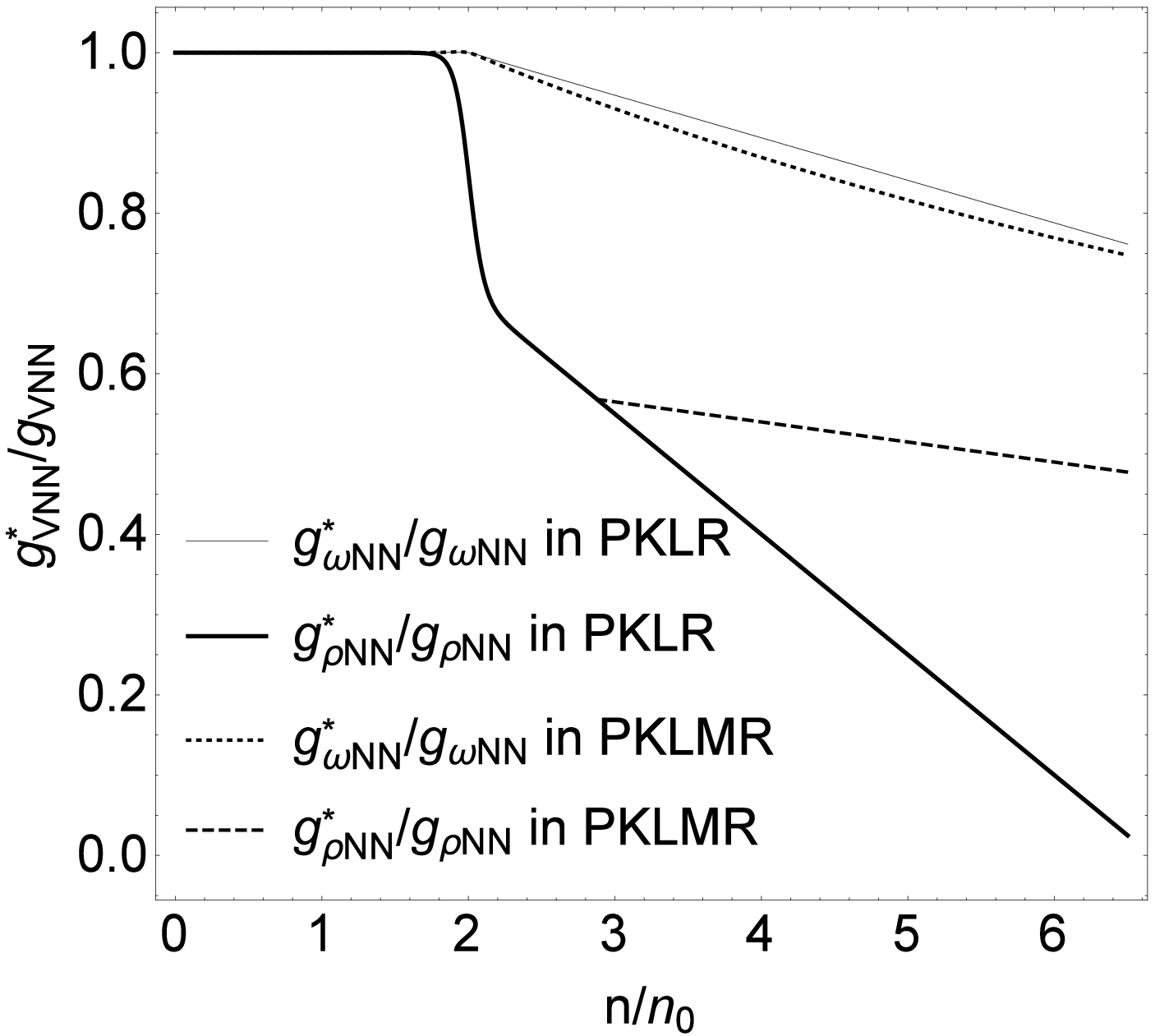}
\caption{ The density dependence of the masses(left) and couplings(right) of vector mesons are compared with each other in PKLR\cite{PKLR} and PKLMR(the present work).
 }\label{mVgV}
\end{center}
\end{figure}

\subsection{Going from R-I ($n \lsim n_{1/2}$) to R-II  ($n > n_{1/2})$}
Of the parameters that undergo changes as density goes across the topology change density, the most striking quantity is the {\it effective} hidden gauge coupling $g_\rho^\ast$.  { What the {\it effective} HLS coupling means in the calculation of EoS depends upon how the Lagrangian $bs$HLS is treated.

In the mean-field approach adopted in \cite{Paeng:2013xya}, it represents $g (g_{V\rho}-1)$ where $g$ is the hidden gauge coupling defined by the matching to QCD at the matching scale and  $(g_{V\rho}-1)$ is an induced factor that is meant to take into account the effect involved in the scale change from the matching scale to the scale from which the mean field is taken. In the spirit of Walecka-type relativistic mean field (RMF) approach implemented with scaling parameters~\cite{Song:2000cu}, corrections to the mean field should be suppressed in the sense that $\tilde{\Lambda}\equiv \Lambda - k_F \rightarrow 0$ as in Landau-Fermi liquid fixed point theory. In other words, it corresponds to the single-decimation RG procedure.

As for the $V_{lowk}$RG that involves the double-decimation strategy,
%
%
we start with the assumption that the nuclear matter at equilibrium can be described in terms of Wilsonian RG~\cite{shankar}. There the quasiparticle interactions are to have vanishing $\beta$ functions in the limit $N\equiv k_F/(\Lambda-k_F)\to \infty$ (where $\Lambda$ is the cutoff for decimation). With the vector mesons and the dilaton of $bs$HLS Lagrangian, considered heavy compared with the Fermi sea scale, integrated out to give the marginal four-point quasiparticle interactions, nuclear matter can be considered to be at its Fermi-liquid fixed point~\cite{Song:2000cu}.  In going beyond the equilibrium density, we continue to assume the Fermi-liquid structure applies.   In Appendix, we will present an admittedly simplistic argument based on skyrmion-crystal matter how a mean-field structure arises in the half-skyrmion phase, which leads -- in the premise that we have adopted -- to a Fermi-liquid structure. Now consider the parameter space of $bs$HLS on top of the Fermi-liquid fixed point. Approaching the IR fixed point with the scale parameter $\tilde{\Lambda}\equiv \Lambda - k_F \rightarrow 0$, the parameters of the EFT Lagrangian should scale such that the $\beta$ function for the quasiparticle interactions tends towards zero  at a given Fermi-momentum $k_F$. Suppose the density is changed from $k_{F1}$ to $k_{F2}$, then certain parameters should change, say the quasiparticle mass as an example, from $m^\ast(k_{F1},\,\tilde{\Lambda}=0)$ to $m^\ast(k_{F2},\,\tilde{\Lambda}=0)$ to preserve $\beta(k_{F1},\, \tilde{\Lambda}=0) = \beta(k_{F2},\, \tilde{\Lambda}=0) = 0$.  This means that the Fermi-liquid fixed point quantities are closely related to each other at  a given density so that $g_V^\ast(k_F,\,\tilde{\Lambda}=0)$ as well as $m^\ast(k_F,\,\tilde{\Lambda}=0)$ should be dependent on $\langle \chi \rangle^\ast$ and $k_F$ to have $\beta(k_F,\,\tilde{\Lambda}=0)=0$.
Thus in the density regime $n \lsim n_{1/2}$ (R-I), the condensate $\langle \chi \rangle^\ast$ given in the mean field~\cite{Paeng:2013xya}, locked to the quark condensate  $\langle\bar{q}q\rangle$, decreases as observed in experiments~\cite{yamazaki}. The hidden gauge coupling remains unscaling in this density regime.

Now going to $n > n_{1/2}$,  the dilaton condensate $\langle \chi \rangle^\ast$ should stay constant as predicted by the theory, i.e., (\ref{chiII}). This requires that $g_{\rho,\omega}^\ast(k_F,\,\tilde{\Lambda}=0)$ scale to preserve  $\beta(k_F,\, \tilde{\Lambda}=0)=0$ as density increases. Then the flow to the VM fixed point for the vector mesons and the dilaton-limit fixed point for the scalar, involving an intricate interplay between their couplings to baryons to preserve  the Fermi-liquid fixed point structure, leads to  the change in density dependence of $g_\rho^\ast$ from $g_\rho^\ast \approx g_\rho$ near $n=n_0$ to $g_\rho^\ast \rightarrow 0$ near the VM fixed point. This changeover is summarized in Fig.~\ref{mVgV}. We admit that although consistent with what we find in the applications reported below, it is non-rigorous and requires a more transparent analytic demonstration.}

\subsection{Toward $\la\theta_\mu^\mu\ra\propto \kappa^4$ for $n > n_{1/2}$}
\label{TMETargue}
Following the argument given in Section \ref{mapping},  we take the results of the skyrmion crystal using $s$HLS Lagrangian, i.e., {\it Skyrmion}$_{\rm sHLS}$, to be equivalent to the mean-field results of $bs$HLS  {\it with the IDDs properly taken into account}.  Now from (\ref{kappa}) of {\it Skyrmion}$_{\rm sHLS}$, we have for $n > n_{1/2}$
\be
\la\chi\ra^\ast\propto \kappa
\ee
which is independent of density. Then, it follows from (\ref{TEMT1}) that
\be
\la\theta_\mu^\mu\ra\propto \kappa^4
\ee
also independent of density.
As will be explained explicitly, the constant TEMT implies that the sound velocity of the dense medium is $v_s^2/c^2=1/3$. This sound velocity is usually associated with the conformal limit, which is arrived at when the TEMT is equal to zero. Here the same velocity is given by a system where TEMT is not equal to zero. Note that $\kappa\to 0$ corresponds to the dilaton-limit fixed point (DLFP). We will discuss how this limit can be arrived at in the framework of {\it EFT}$_{\rm bsHLS}$.

The question is: Can the $V_{lowk}$ RG treatment give the same sound velocity $1/\sqrt{3}$?

To address this question, we recall that the $bs$HLS at the mean field corresponds to Landau Fermi-liquid fixed point with the vanishing $\beta$ functions for the  Landau quasi-particle interactions and effective quasiparticle mass $m^\ast$ for fixed $k_F$. In our terminology, this is precisely the first RG decimation in the $V_{lowk}$ RG. Now with our main assumption that the $\beta$ functions remain 0  as $(\Lambda-k_F)/k_F\to 0$,  we should expect that the second decimation going beyond the Fermi-liquid fixed point -- with loop corrections, e.g, ring diagrams -- should leave the TEMT unmodified. This means, following from the skyrmion-crystal result, that the TEMT should be a constant independent of density in the leading order in chiral-scale symmetry~\cite{LMR-2016}. The caveat is that in Nature, both symmetries, intricately locked to each other, are explicitly broken. Therefore we cannot expect that the TEMT will be exactly density-independent.

What we aim then is to adjust the {\it only}  IDD property in the region $n > n_{1/2}$ that is unconstrained in {\it Skyrmion}$_{\rm sHLS}$, namely the approach to the VM fixed point of the hidden gauge coupling $g_\rho^\ast$.  It turns out that what is needed is to have the effective $g_\rho^\ast$ drop less rapidly than in \cite{PKLR} slightly after $n_{1/2}$ and have the VM fixed point reach at  a higher density, say, $n >  25 n_0$ than $\sim 7n_0$, which was taken in \cite{PKLR}.  This postponement to higher density for a phase transition, be that chiral restoration coinciding with the VM fixed point or deconfinement, the precise value of which is unknown in the given theoretical framework and also in QCD proper -- so it is totally arbitrary  -- simply implies that there will be no phase transitions in the range of densities relevant to compact stars.
\begin{figure}[h]
\begin{center}
\includegraphics[width=10.0cm]{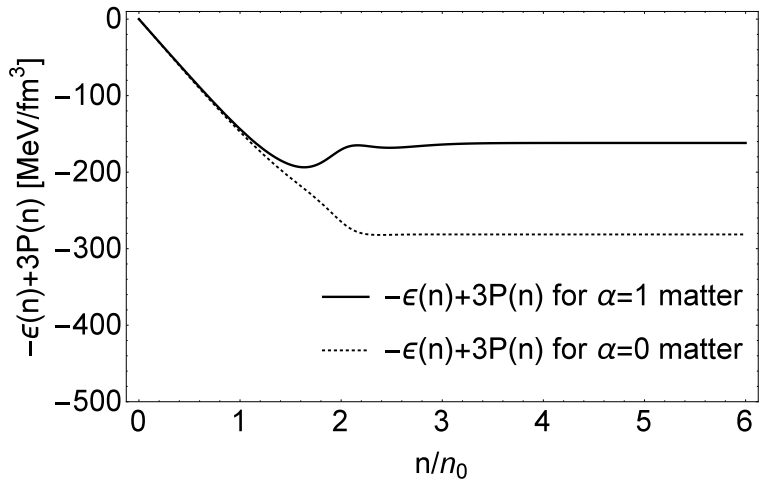}
\caption{ $-\epsilon(n) + 3P(n)$ vs. density for $\alpha=0$ (nuclear matter) and $\alpha=1$ (neutron matter)
 }\label{TEMT}
 \end{center}
\end{figure}

The required adjustment for the IDD for the $g_\rho^\ast$ to get a constant density-independent TEMT for $n > n_{1/2}$ is summarized in Fig.~\ref{mVgV}. {The density dependence of $\frac{g_{\rho NN}^\ast}{g_{\rho NN}}$ is given by\footnote{ The precise form of the scaling has no significance. What matters is the two changes in slope seen in Fig.~\ref{mVgV}, one at $ n_{1/2}\sim 2n_0$ and another at $\sim 3n_0$. They could be smoothed and modified with refinement in the renormalization group treatment, which would alleviate the flattening in the scaling. In fact it has been shown that in a simple chiral model of interacting mesons and nucleons, a functional renormalization group (FRG) method -- which is nonperturbative improvement over standard chiral perturbation theory and mean-field approximations -- flattens appreciably the dropping in density of the chiral order parameter at densities exceeding $n_0$~\cite{FRG-Weise}.  It is not unlikely that this method contains more than what's included in our Fermi-liquid approach via $V_{lowk}$RG and could modify the {\it effective} IDDs for the hidden gauge couplings.}
 \begin{equation}
 \frac{g_{\rho NN}^\ast}{g_{\rho NN}} = \left\{ \begin{array}{cc} 1 - 0.15 \frac{n}{n_0}\quad {\rm for} \quad n_{1/2} < n < 2.88n_0 \\ 0.568 - 0.025\frac{(n - 2.88n_0)}{n_0} \quad {\rm for} \quad 2.88n_0 < n < n_c  \end{array} \right.
 \end{equation} which gives $\frac{m_{\rho }^\ast}{m_{\rho }} = \frac{g_{\rho NN}^\ast}{g_{\rho NN}} = 0$ at $n = 25.6 n_0$. }
The resulting TEMT is plotted in Fig.~\ref{TEMT}.
It turns out, quite surprisingly, that this result can be captured  very well by the simple formula for $n\geq n_{1/2}$
\begin{equation}
E_0/A = -m_N + B \left( \frac{n}{n_0} \right)^{1/3} + D\left( \frac{n}{n_0} \right)^{-1} \label{EoSRII}
\end{equation} which is an analytic solution of $\frac{d\, P}{dn} = \frac{1}{3} \frac{d\,\epsilon}{dn}$ which assumes the density independent TEMT in $n > n_{1/2}$.
To confirm that it is a good parameterization, it is checked with the  energy per particle $E_0/A$ and the pressure $P$ of both the symmetric nuclear matter and pure neutron matter as well as the
symmetry energy $E_{\rm sym}$ computed in our $V_{lowk}$RG. The fit is plotted in Fig.~\ref{E0A}. The fit parameters are
\be
B_{\alpha=(0,1)}&=& (570 \textnormal{ MeV},\, 686 \textnormal{ MeV}), \\
D_{\alpha=(0,1)}&=& (440 \textnormal{ MeV},\, 253\textnormal{ MeV}).
\ee
\begin{figure}[h]
\begin{center}
\includegraphics[width=7.5cm]{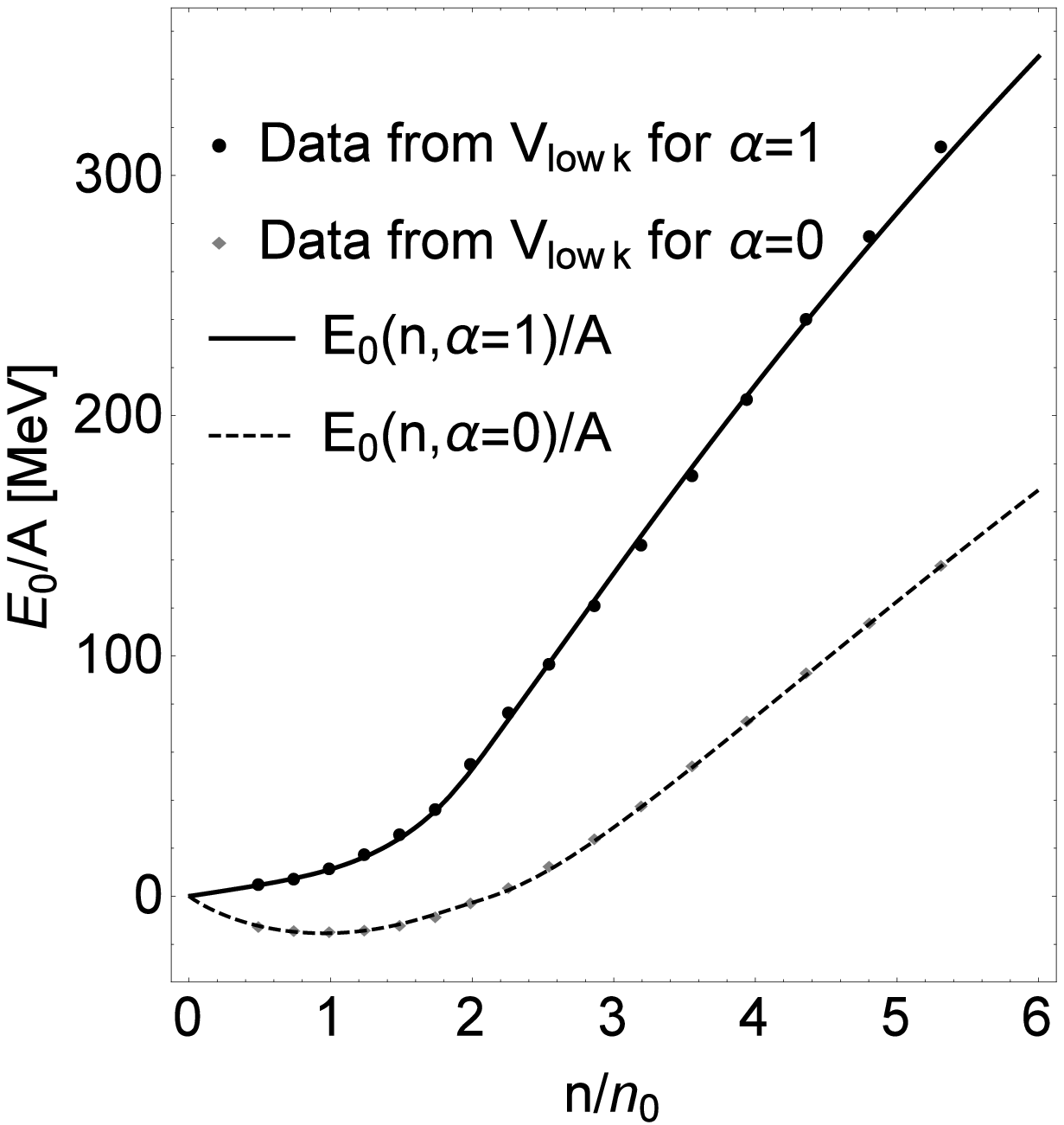}\includegraphics[width=7.5cm]{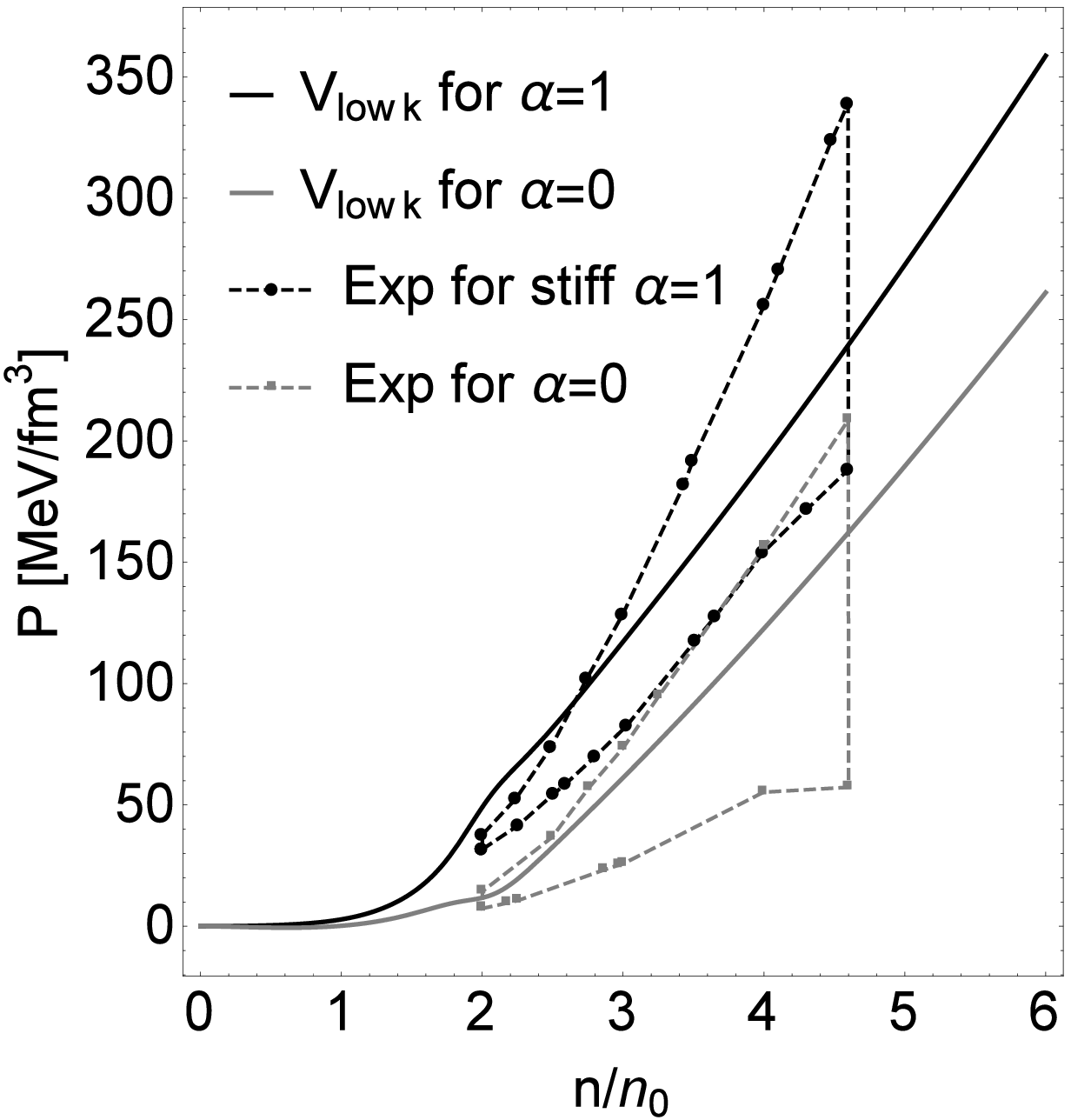}
\caption{ $E_0/A$(left) and  $P$ (pressure) (right) for $\alpha =1$ and $\alpha = 0$ matter. The empirical values(named ``Exp'') for the pressure are taken from Danielewicz\cite{Danielewicz:2002pu}.
 }\label{E0A}
 \end{center}
\end{figure}

It is seen that the results of the $V_{low\,k}$-RG with $bs$HLS are well reproduced for both the neutron matter($\alpha=1$) and the symmetric nuclear matter($\alpha =0$) by the formula (\ref{EoSRII}). There are small deviations at high density, for instance in $E_{\rm sym}$, but these could be ameliorated by a refined form for the IDD in the bare parameters. What transpires is that the simple form is good enough to confirm that our scenario is well captured in $V_{low\,k}RG$ with $bs$HLS and to correctly give the scaling property of the $bs$HLS parameters essential for the TEMT in the high-density regime $n> n_{1/2}$. As we see in Fig. \ref{E0A}, the EoS's for both the $\alpha=1$ and $\alpha=0$ matter are stiff in $n> n_{1/2}$. This feature is consistent with the available heavy ion data. The dotted and dashed lines in Fig. \ref{E0A} and \ref{Esymh} depicting the pressure and the symmetry energy represent the empirical constraints for the EoS coming from heavy-ion data given by Danielewicz\cite{Danielewicz:2002pu} for the pressure, Li et al\cite{Li:2005jy} and Tsang et al\cite{Tsang:2008fd} for the symmetry energy respectively. Note that the cusp seen in the skyrmion crystal simulation is smoothed in the $V_{lowk}$ RG into a soft-to-hard EoS at $n_{1/2}$ as observed, e.g.,in \cite{PKLR}.
\begin{figure}[h]
\begin{center}
\includegraphics[width=7.5cm]{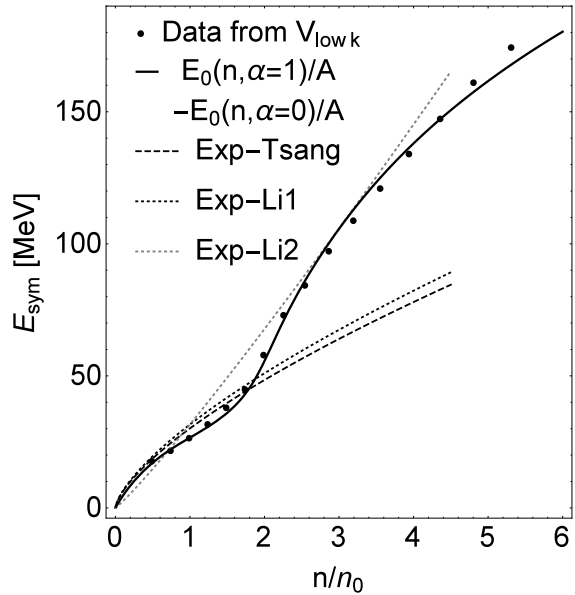}
\caption{ The symmetry energy $E_{sym}$  vs. the density is shown. The empirical values(named ``Exp'') for the symmetry energy are taken from Li et al\cite{Li:2005jy} and Tsang et al\cite{Tsang:2008fd}.
 }\label{Esymh}
 \end{center}
\end{figure}

\subsection{Predictions for massive compact stars}
Given the EoS described above,  it is a standard procedure to use the Tolman-Oppenheimer-Volkoff(TOV) equation to calculate the properties of compact stars.
The leptons that participate in beta stability need to be taken into account as they determine the proton fraction $x\equiv n_p/n =Z/(N+Z)$.  Beta equilibrium involving leptons, i.e., electrons and muons, $n\leftrightarrow p+e^- + \bar{\nu}_e$ and $n\leftrightarrow p+\mu^- + \bar{\nu}_\mu$ and charge neutrality imply
\be
n_p=n_e+n_\mu,\ \ \mu_n=\mu_p+\mu_e,\ \ \mu_e=\mu_\mu
\ee
where
\be
\mu_{n,p}=\Big(\frac{\partial \epsilon}{\partial n_{n,p}}\Big)_V.
\ee
It is a good approximation to assume $\mu_l$ to be the chemical potentials of those of free Fermi gas of electrons and muons. The proton fraction $x\equiv n_p/n$ for matter in beta equilibrium is then determined by minimizing, for a given nucleon density $n$, the total energy per particle $E_0/A$ (in Fig.~\ref{E0A}) plus the contributions from leptons and from the rest mass of the nucleons. Here, we take $\mu_n - \mu_p \approx 4E_{\rm sym}(n) \alpha$. The resulting proton fraction is given in Fig.~\ref{protonfraction}.
\begin{figure}[h]
\begin{center}
\includegraphics[width=8cm]{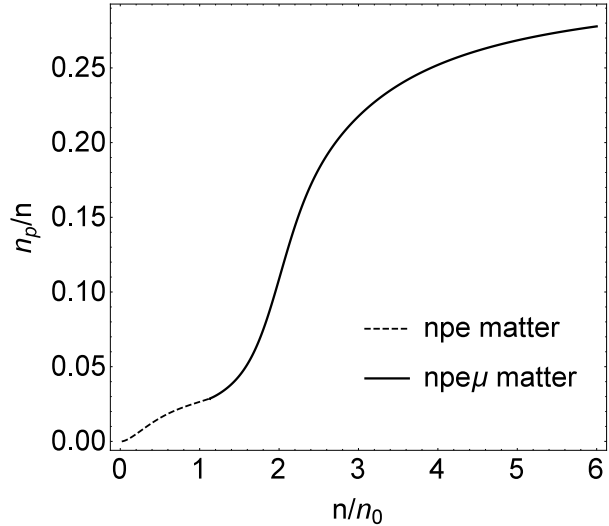}
\caption{The proton fraction $x\equiv n_p/n$ in neutron star matter with beta equilibrium. ``npe matter'' and ``npe$\mu$ matter'' are composed of neutron, proton, electron and neutron, proton, electron, muon respectively.
 }\label{protonfraction}
 \end{center}
 \end{figure}
 It is appropriate to mention at this point that the predicted proton fraction is such that at $n\sim 2.5n_0$, it exceeds the threshold density for the direct URCA process for star cooling, $x_{dURCA}\approx 0.14$. It also affects significantly the maximum star mass, say, about 10\% in comparison with the EoS of pure neutron matter.  We will comment on this matter in Section 6.

The same EoS that gives the constant TEMT, Fig.~\ref{TEMT}, with the beta equilibration suitably included,   gives the star mass vs. radius and the star mass vs. central density, as shown in Fig.~\ref{mass-radius}.
The result is consistent with the well-measured value  $M=2.01 \pm 0.04 \,M_\odot$\cite{Antoniadis:2013pzd}. The radius for this mass object is not yet pinned down but what we get is in the ranges discussed in the literature. It is reasonable to conclude that we have here an evidently respectable EoS for massive compact stars. Up to the predicted central density $\sim 5.1n_0$, there are no other degrees of freedom than the pNG bosons ($\pi$ and $\sigma$) and vector mesons (with or without explicit baryons). There are neither hyperons nor quarks. There is no de-confinement intervening in the structure.
\begin{figure}[h]
\begin{center}
\includegraphics[height=6.cm]{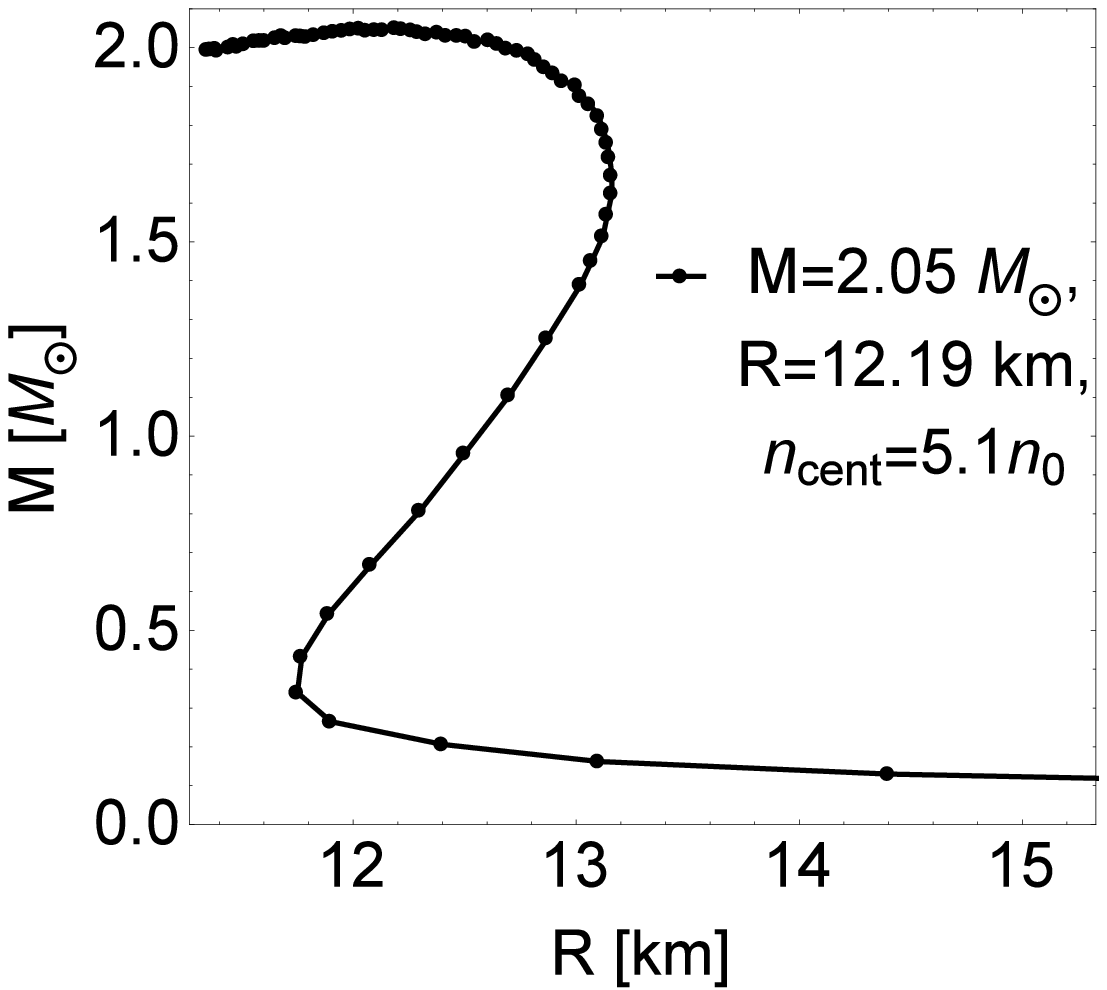}\includegraphics[height=6cm]{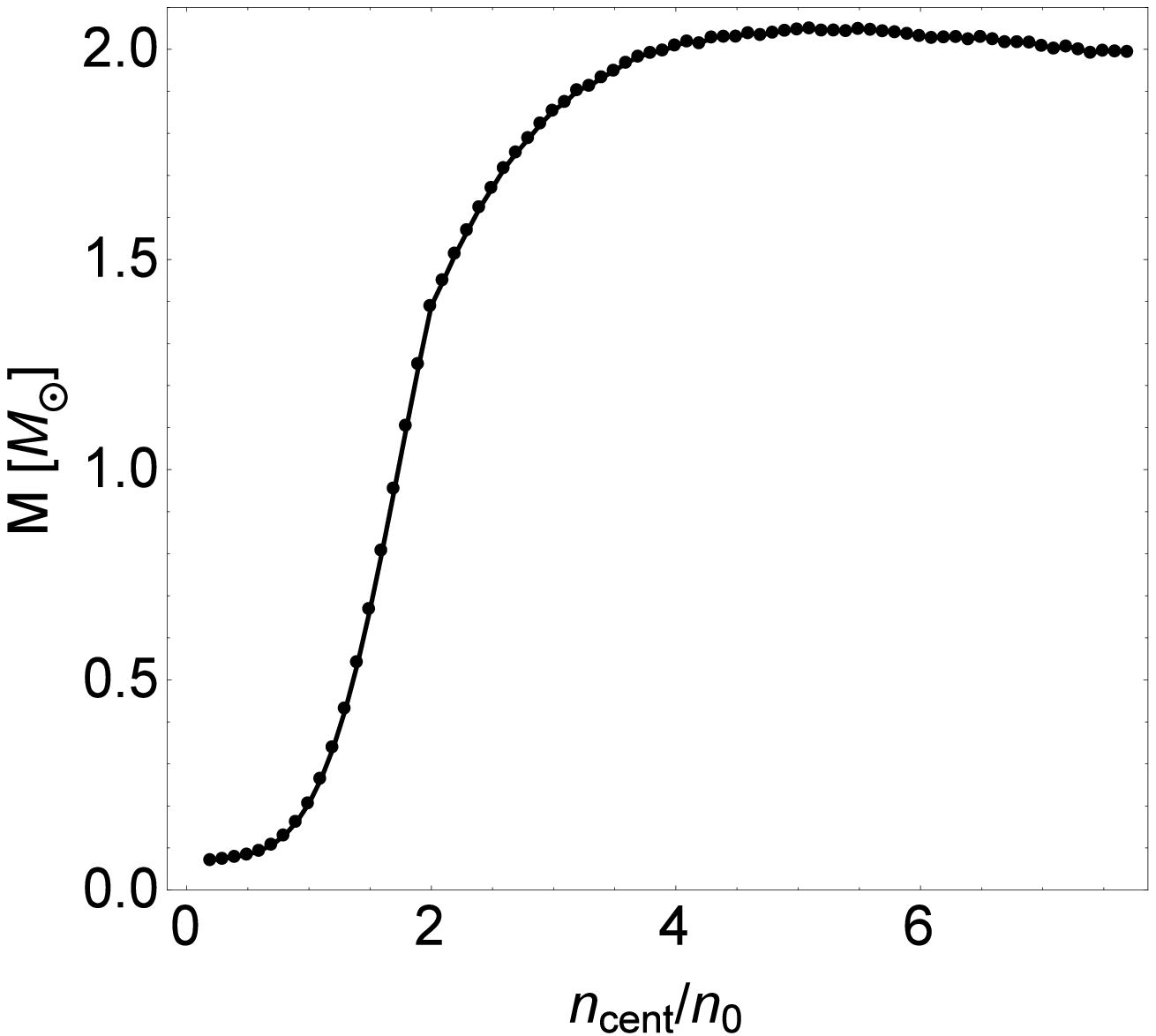}
\caption{ Mass $M$  vs. radius $R$ and $M$ vs. central density $n_{\rm cent}$.
 }\label{mass-radius}
 \end{center}
\end{figure}

Now what about the sound velocity for the $\sim 2$ solar mass neutron star? Here is a surprising result.

The trace of energy-momentum tensor and the sound velocity are related by
\be
\frac{\partial}{\partial n} \la\theta_\mu^\mu\ra=\frac{\partial \epsilon (n)}{\partial n} \Big(1-3\frac{v_s^2}{c^2}\Big)\label{derivTEMT}
\ee
with $v_s^2/c^2=\frac{\partial P(n)}{\partial n}/\frac{\partial \epsilon(n)}{\partial n}$. Since from Fig.~\ref{TEMT}, we have the TEMT (nearly) independent of density, the left-hand side of (\ref{derivTEMT}) is (nearly) zero. Assuming that there is no extremum in the energy density in compact star matter, then $\frac{\partial \epsilon (n)}{\partial n} \neq 0$. It therefore follows that
\be
v_s^2/c^2\approx 1/3.
\ee
The prediction of our approach is given in Fig.~\ref{sound}. Our EoS, while it gives quite different sound velocities at densities below $n\sim 3n_0$ for the $\alpha=0$ and $\alpha=1$ matters, makes both of their velocities approach rapidly $\sqrt{1/3}$ for $n\gsim 3n_0$. Note that in the range of densities considered here, $\la\theta_\mu^\mu\ra\neq 0$, so scale symmetry is not restored. We identify this phenomenon as a precursor to the emergence of scale symmetry in dense medium.

This is one of our principal findings, which could be considered as a  unique prediction of the theory. What this implies vis-\`a-vis with the EoS is discussed in the discussion section.
\begin{figure}[h]
\begin{center}
\includegraphics[height=7cm]{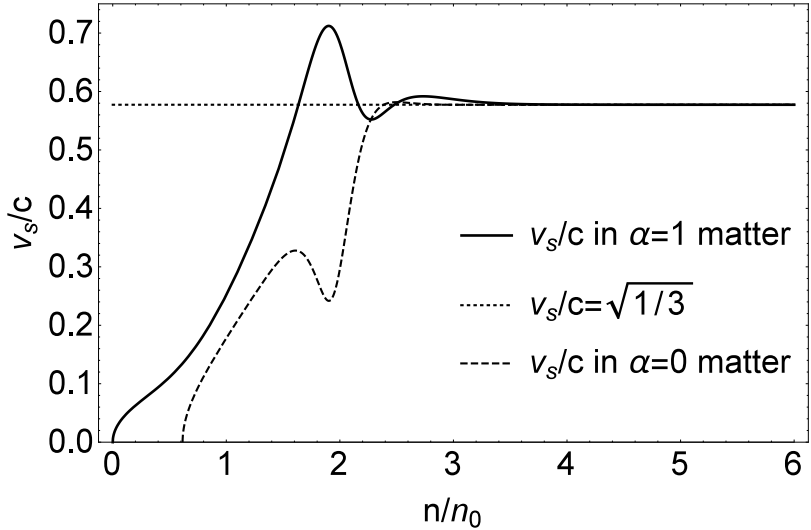}
\caption{ Sound velocity vs. density.
 }\label{sound}
 \end{center}
\end{figure}
\subsection{Gravity wave and tidal deformability}

Our $V_{lowk}$-RG approach makes certain predictions that differ from  the  phenomenological approaches found in the literature, in some cases, rather strikingly such as the sound velocity, the density-independent trace of energy momentum tensor and emerging symmetries invisible in QCD in the vacuum. One asks whether these predictions can be distinguishable from other models and can be seen in measurable quantities. Phenomenological models with a large number of parameters can be adjusted to fit most, if not all, of what's available from astrophysical observables, so what is needed is a pristine signal for the unorthodox predictions made in the present approach. For this purpose, we look at what could be seen in gravity waves emitted in the coalescing of neutron stars.  We discuss as a specific case the tidal deformability predicted by the EoS given by $V_{lowk}$-RG and compare it with what's available in the literature.

With the EoS given by $bs$HLS above, we could calculate the tidal deformability $\lambda_D$ as shown in the left panel of Fig. \ref{lambda} and also the dimensionless quantity $\Lambda_D = \lambda_D G \left( \frac{c^2}{G\,M_\odot} \right)^5 \left( \frac{M_\odot}{M}\right)^5$ with the gravitational constant $G$.
\begin{figure}[h]
\begin{center}
\includegraphics[width=7.5cm]{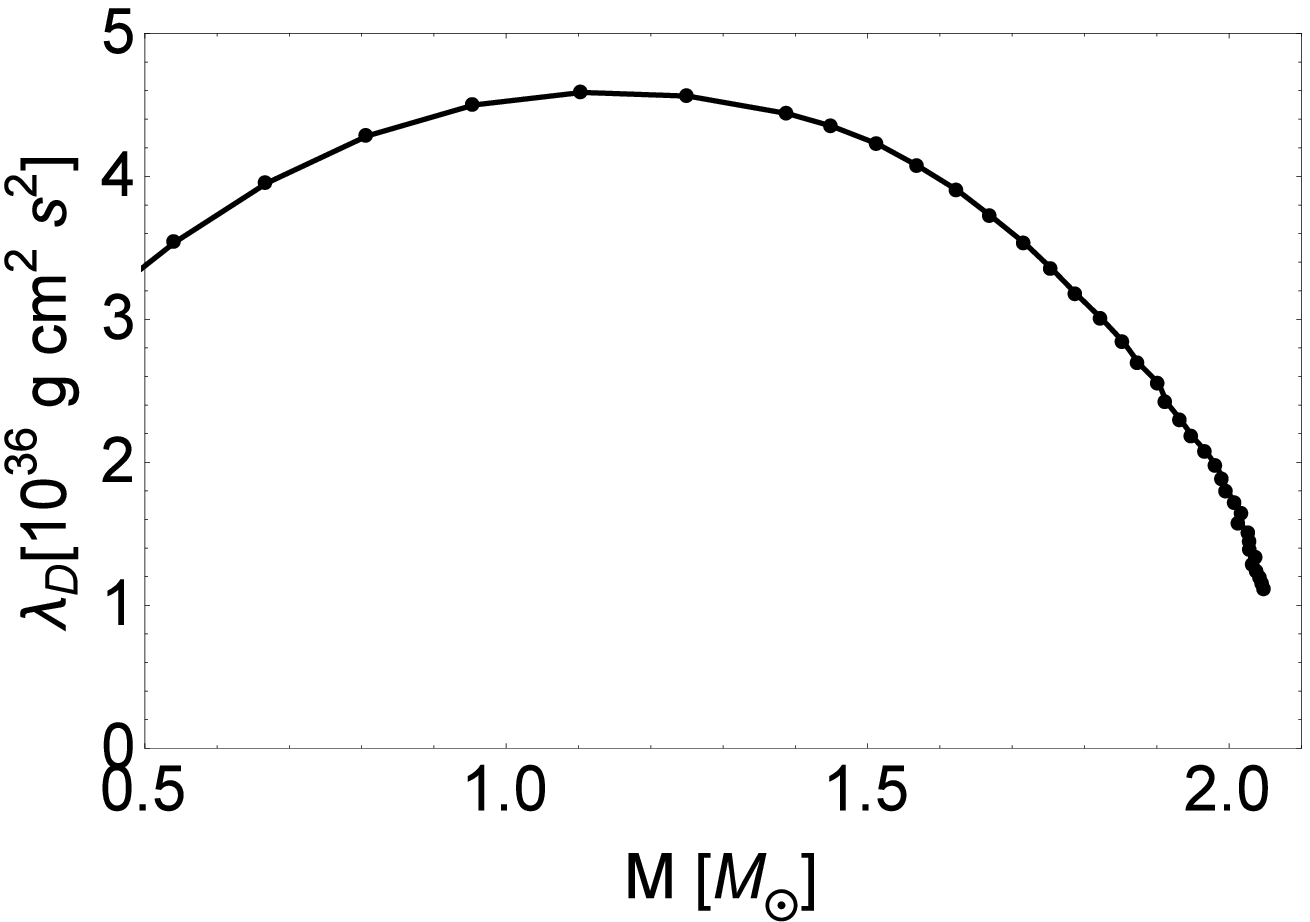}\includegraphics[width=7.5cm]{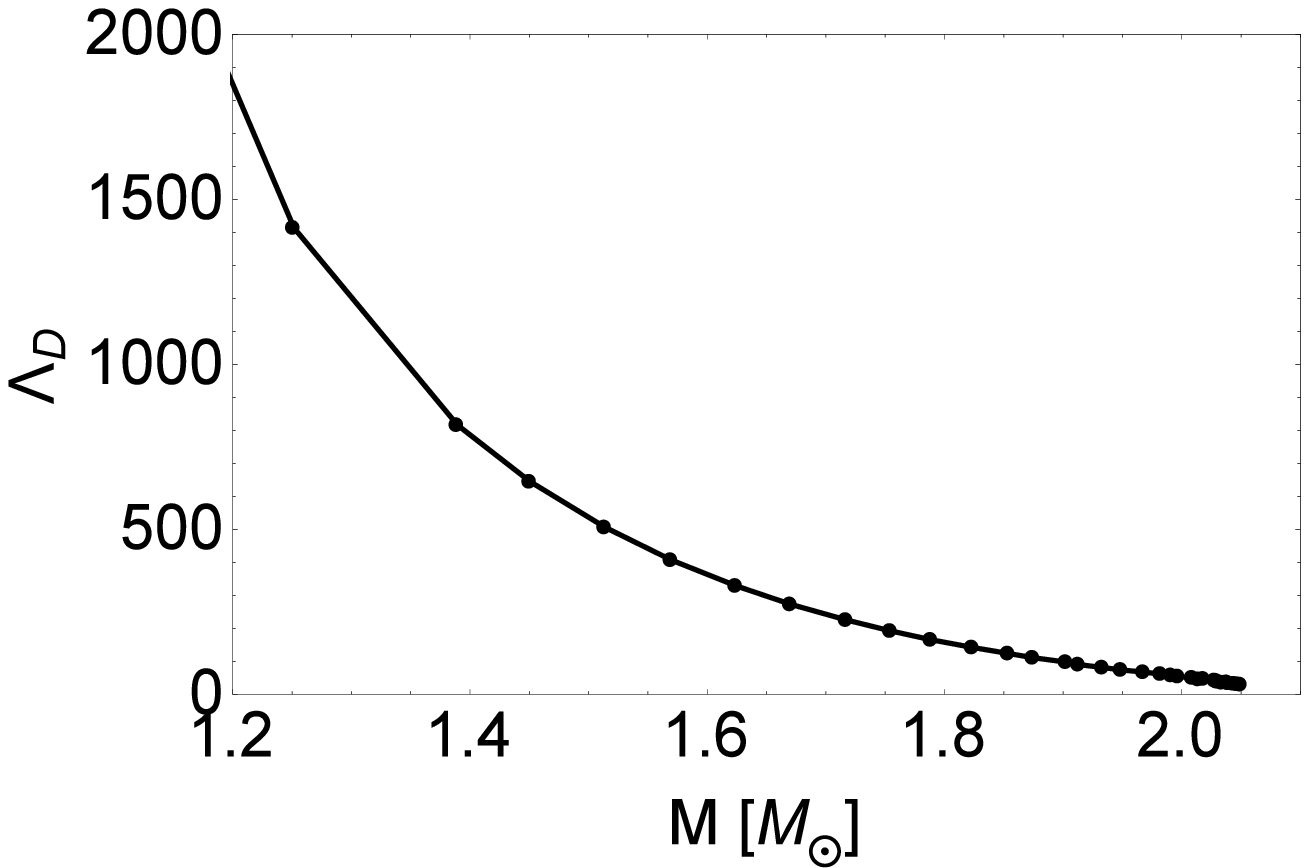}
\caption{ The deformability $\lambda_D$($\Lambda_D$) vs. the neutron star masses are shown.
 }\label{lambda}
 \end{center}
\end{figure}

The most interesting mass range is $1.3 -1.5 M_{\odot}$, for which most of neutron star masses so far discovered are populated.  One may expect more abundant gravitational wave emissions from binary collapses with this mass range than from the binaries with different  masses. The deformation parameters $\lambda_D$ and dimensionless parameters $\Lambda_D$ for mass ranges $1.1 -1.5 M_{\odot}$  are calculated  in Table \ref{tidaldef} with the corresponding radii and central densities.
The deformability parameter  for $\sim 1.4M_{\odot}$ is found to be   $4.44 $ in unit of $10^{36}{\rm g cm^2 s^2}$, which  can be compared with those of different EoS's:  For example, the EoS's of SLy\cite{DH}, AP3\cite{AP3} and  MPA1\cite{MPA}  for the same mass give $\lambda_D = 1.70, ~2.22$ and $2.79$ respectively~\cite{PRD/81/123016}.

\begin{table}
\begin{center}
\begin{minipage}{.8\textwidth}
\caption{Tidal  deformabilities  }
\label{tidaldef}
\begin{center}
\begin{tabular}{c c c c c}
\hline
\hline
$M/M_{\odot}$ & $n_c/n_0$   &  $\lambda_D/(10^{36}{\rm gcm^2s^2})$   & $\Lambda_D/100$  &   $ R/{\rm km}$ \\
\hline
$1.10$   & $1.8$ & $4.59$ & $26.5$ & $12.7$   \\
$1.25$   & $1.9$ & $4.56$ & $14.1$ & $12.9$  \\
$1.39$   & $2.0$ & $4.44$ & $8.15$ & $13.0$  \\
$1.45$   & $2.1$ & $4.35$ & $6.44$ & $13.0$  \\
$1.51$   & $2.2$ & $4.22$ & $5.06$ & $13.1$  \\
\hline
\hline
\end{tabular}
\end{center}
\end{minipage}
\end{center}
\end{table}
 On the other hand,  in the  higher mass range near to $\sim 2 M_{\odot}$, the deformability parameter in this work is found to be   not much different from   those EoS above mentioned.

For the measurability analysis of deformability from gravitational waves, it is better  to use the  dimensionless form of deformability parameter, since the deviation due to the neutron star deformation from point particle approximation turns out to be  expressed by $\Lambda_D$ rather than $\lambda_D$ itself.
 From Table \ref{tidaldef}, one can see that the dimensionless parameter $\Lambda_D = 815$ for $M=1.4M_{\odot}$ predicted in our approach is  much larger than  $\Lambda_D = 312, 408, 512$ for  those EoS  of SLy\cite{DH}, AP3\cite{AP3} and  MPA1\cite{MPA} respectively. The differences between those and this work  are  $\delta \Lambda_D = 503, 407, 303$.   It is interesting to note that recent numerical analysis\cite{read, HKSS} demonstrated the  measurability  of tidal deformations  determined by the change of late inspiral wave forms for $\delta \Lambda_D > 100$.  This  implies that the differences between this work  and  those above mentioned,   which are larger than the distinguishability criteria \cite{HKSS}, are expected to be measurable in the forthcoming observations at aLIGO and aVirgo.  It should be mentioned that the central densities of these neutron stars, $1.1-1.5M_{\odot}$, are not  high enough to probe the half-skyrmion phase, $n_c \leq n_{1/2}(=2n_0)$. Most of these neutron star interiors up to the central core is governed by the EoS for  $n<2n_0$:
for a neutron matter $E/A (MeV) = 9.11~u + 2.14~u^{4.08}$ and $E/A(MeV) = -45.5~u + 30.1~u^{1.54}$ for a symmetric nuclear matter.  The major characteristics  of this work is the $V_{lowk}$-RG approach,  which is  quite  different from others in this density region, such as  giving much bigger  deformation parameters  than others.

For  large mass neutron stars, the  central density is higher than the threshold for  the half skyrmion phase, in which TEMT becomes  independent of density with sound velocity $v^2_s = 1/3$:  $E/A(MeV) = -940 + 253/u + 686~u^{0.33}$ and  $E/A(MeV) =-940 + 440/u + 570~u^{0.33}$ for a neutron matter and symmetric matter respectively.
In this mass range, as the mass is increasing  the deformation parameter decreases and  near $2M_{\odot}$  the deformation parameter  becomes not much different from others as seen in \cite{KLL}.  The distinguishability  of EoS by the tidal deformation using gravitational waves does not seem to be  effective for these high mass binaries.   Hence  there may be scant  possibility  for  the tidal deformability  to  determine whether our scenario with the sound velocity 1/3 is distinguishable from other garden-variety models  in the gravitational wave forms from the higher mass neutron star binaries.  However  our scenario is quite different from that of others of phenomenological approaches where the sound velocity is typically $> 0.5$ which would make them in tension with the conformality bound of Bedaque and Steiner\cite{bedaque}.

Another possibility of probing  EoS at  higher  density is the gravitational waves emitted just after merger.   When they start merging  after inspiral and   the density  of  the merger remnant of the colliding matter becomes  much  higher  than the core of the original stars.   That  is,  at the intermediate  stage before becoming a black hole,    1.5 solar mass binaries(even though the core density is not higher than $2n_0$ )  can make a merger remnant  of  higher density up to  $\sim 5.5 n_0$ \cite{radice}. The gravitational wave  forms  during merger  of course  are then  expected to carry the information of  EoS at the higher density. From the numerical simulations, it is known that the relevant frequency range of GW during merger is much higher than inspiral period,  more than kilo Herz.   Recent analysis\cite{radice} demonstrates that the EoS softening at higher density encoded in gravitational wave amplitudes can be  detectable up to distances of $\sim  20$  Mpc with advanced detectors(aLIGO and aVirgo) and $\sim  150$ Mpc with third generation detectors(for example, Einstein Telescope(ET)).  This shows an additional interesting window in gravitational wave observations where the high density hadronic matter can be probed  to be able to distinguish our scenario from others.

\subsection{Going to the DLFP}

As emphasized, there is no reason to adhere to the possibility that the sound velocity of compact stars is exactly $1/\sqrt{3}$.  This is because the TEMT cannot be exactly constant of density in the density range of compact stars in nature, not exceeding much beyond the central density $\sim 5 n_0$. Among others, there is explicit symmetry breaking, chiral as well as scale, that needs to be accounted for.  However at some higher density, say, $\gsim  7 n_0$, the dilaton-limit fixed point with $\la\chi\ra^\ast\to 0$ may be approached if not reached exactly on top. In this case then the sound velocity will approach nearly exactly $1/\sqrt{3}$.  This would correspond to the emergence of scale symmetry in dense medium even though  $\beta^\prime$ is not equal to zero.

We have no rigorous argument for  the existence of such a phenomenon. However we can entertain a conjecture that could be validated with better understanding of the  structure of the theory.

Given the ``walking" dilaton condensate for $n>n_{1/2}$, the question is how one can induce $\langle \chi \rangle^\ast \rightarrow 0$ to reach the DLFP in dense matter.  This may appear difficult to answer in an affirmative way. However if  one accepts the density dependence of $g_\omega^\ast$ as prescribed in our theory,  with a  change after $n = n_{1/2}$,  the following is a possible scenario for the transition to the DLFP.

When $g_\omega^\ast$ decreases, albeit slowly, as density increases in the density regime $n\gsim n_{1/2}$, the $\omega$-repulsion will get reduced but if it still counter-balances the $\sigma$-attraction enough as density goes above $n_{1/2}$, then the matter will remain stable. But, if the $g_\omega^\ast$ becomes further suppressed so that the $\omega$-repulsion becomes comparable with or weaker than the $\sigma$-attraction, then the matter will become unstable at increasing density {\it unless} the parameters in $bs$HLS are modified.
Approaching the dilaton-limit fixed-point density $n_{DLFP}$, the matter could be stabilized by intricate parameter changes such that  $\langle \chi \rangle^\ast \rightarrow 0$ with the quasiparticles decoupled from the $\rho$ and $\omega$ and the dilaton becoming massless $m_\sigma^\ast\propto  \langle \chi \rangle^\ast\rightarrow  0$.

We can offer a heuristic mean-field argument that supports
 the scenario discussed above, namely that the walking $\langle \chi \rangle \sim $ constant induces the system to go to DLFP.  In the mean-field approximation with $bs$HLS, the $\omega$-NN interaction could be reinterpreted to include the $\omega$-exchange force between the nucleons given by loop contributions to the $\omega$-NN vertex, such as  Fock terms. If $g_\omega^\ast$ drops at increasing density, then the $\omega$-repulsion will get reduced. This density dependence of the $\omega NN$-repulsion should be taken into account in doing the mean-field calculation. For simplicity, we include the effects from the $\omega$ exchange and higher order terms of $bs$HLS for the $\omega NN$ repulsion into the density dependence of $g_{V\omega}^\ast$ following \cite{Song:2000cu,Paeng:2013xya}. Then the thermodynamic potential with a density-dependent $g_{V\omega}^\ast$ and $g_\omega^\ast$ for $bs$HLS can be written as
\begin{eqnarray}
\Omega(\chi,\, n)
&=&
\frac{1}{4\pi^{2}} \left[ 2 E_{F}^{3} k_{F} - m_{N}^{\ast 2} E_{F} k_{F}
{}- m_{N}^{\ast 4} \ln \left( \frac{E_{F} + k_{F} }
{m_{N}^{\ast}} \right) \right]
{} + \frac{\left(g_{V\omega}^\ast -1 \right)^2}
{2f_{\sigma\omega}^2 \langle\chi\rangle^2/f_{\sigma}^2} n^2
\nonumber\\
&&
{} -  \frac{ m_\chi^2}{8} f_{\sigma}^{2}
\left\{\left( \frac{\langle\chi\rangle^2}{f_\sigma^2}\right)^{2}
\left[ \frac{1}{2} - \ln\left(\frac{\langle\chi\rangle^2}{f_\sigma^2}\right)
\right] - \frac{1}{2} \right\} -\mu(n) n\,,
\label{omega1}
\end{eqnarray}
where $E_F = \sqrt{k_F^2 + m_N^{\ast\, 2}}$
and the chemical potential is given as a function of density $n$ by
\begin{equation}
\mu(n)
= E_F(n) + \frac{\left( g_{V\omega}^\ast -1\right)^2}
{f_{\sigma\omega}^2 \langle\chi\rangle^2/ f_\sigma^2} n
{}+ \frac{\left( g_{V\omega}^\ast -1\right)}
{f_{\sigma\omega}^2 \langle\chi\rangle^2/ f_\sigma^2}n^2
\frac{\partial \left( g_{V\omega}^\ast -1\right) }{\partial n}
\end{equation} including the rearrangement term $\frac{\left( g_{V\omega}^\ast -1\right)} {f_{\sigma\omega}^2 \chi^2/ f_\sigma^2}n^2 \frac{\partial \left( g_{V\omega}^\ast -1\right) }{\partial n}$. As shown in \cite{Paeng:2013xya}, if $g_{\omega NN} \sim n^{-\frac{1}{2}}$ so that $\langle \chi \rangle \sim$ constant, the thermodynamic potential (\ref{omega1}) becomes
\begin{equation}
\Omega(n)_{{\rm walking}\, \langle \chi \rangle} \approx - \frac{1}{4\pi^2} \left[ \frac{2}{3} E_F k_F^3 - m_N^{\ast\,2}E_F k_F  + m_N^{\ast\,4} \ln\left( \frac{E_F + k_F}{m_N^\ast}\right) \right] +  V(\langle \chi \rangle) \label{omega2}
\end{equation} which gives $\langle \theta^\mu_\mu \rangle \approx$  constant. This can be  well fitted by $\Omega(n)_{{\rm walking}\, \langle \chi \rangle} \approx -A\times k_F^4 - B < \Omega(n)_{DLFP} = -\frac{1}{6\pi^2}k_F^4$ at low density, where $0< A < \frac{1}{6\pi^2}$, $0 < B$\footnote{By doing the mean-field calculation with $bs$HLS, it is shown that $\langle \chi \rangle = 0$ if $g_{V\omega} =1$ so that $\Omega(n)_{DLFP} = -\frac{1}{6\pi^2}k_F^4$. }. Thus we find that $\Omega(n)_{{\rm walking}\, \langle \chi \rangle}$ becomes greater than $\Omega(n)_{DLFP}$ as density increases above some density.  This means that the baryonic matter with $g_{\omega NN}^\ast = \langle \chi \rangle =0$ at DLFP becomes more favorable energetically  than the matter with a walking $\langle \chi \rangle$. This will trigger a (first-order)  transition to DLFP.  We recall that  $g_{\omega NN} = 0$ is required in RG analyses of the $bs$HLS parameters to arrive at  the DLFP as an IR  fixed point~\cite{Paeng:2013xya} and also for the skyrmion matter simulated on  lattice~\cite{Lee:2003eg,PRV1} to arrive at $\langle \chi \rangle \rightarrow 0$ which makes the energy density be $E/B/V \propto \frac{1}{L^4} \propto n^{4/3}$ so that $\langle \theta^\mu_\mu\rangle = 0$ and $\langle \chi \rangle = 0$. These results support that the density dependence of $g_{\omega NN}^\ast$ related to the behavior of $\langle \chi \rangle$ is important for going to the DLFP. How the $\omega$-nucleon interaction is related to the emergence of the scale symmetry could be studied in the scale-chiral Lagrangian approach formulated in  \cite{MR-omega}.

Furthermore if $n_{DLFP} \sim n_{VM}$, the vector manifestation could set in with $f_{\sigma\rho}^\ast \rightarrow f_\pi^\ast$ and $m_\rho^\ast(\sim g_\rho^\ast) \rightarrow m_\pi^\ast \rightarrow 0$ approaching the VM fixed point, together with the GL satisfied by $f_\sigma^\ast \rightarrow f_\pi^\ast$ and $m_\sigma^\ast \rightarrow m_\pi^\ast \rightarrow 0$ approaching the DLFP. Note that  in this scenario the sound velocity $v_s/c\simeq  1/3$ will have no discontinuity as $\la\chi\ra$ changes from ``walking" to ``running" toward zero. It may be that the DLFP and the VM fixed point coincide, in which case $\pi$, $\sigma$, $\rho$ and $a_1$ could come together into Weinberg's ``mended symmetry"~\cite{weinberg}. Whether this state of matter can be reached in Nature will require a treatment in the half-skyrmion phase which in $s$HLS sets in for $n>n_{1/2}$ which we believe must overlap with the quarkyonic phase in which quark degrees of freedom are to figure~\cite{fukushima}.

We admit that the scenario described in this section is highly speculative.

\section{Remarks}
It is shown in this paper that an EoS constructed with an EFT Lagrangian whose intrinsic density dependence is inherited from QCD combined with what one can extract from a skyrmion structure of dense matter, that give a satisfactory description of currently measured massive compact-star properties, predicts the sound velocity approaching the conformal limit $1/\sqrt{3}$. The essential ingredient was that the trace of the energy-momentum tensor in the chiral limit of QCD approaches, at density $n \gsim 2 n_0$, a density independent constant $\kappa^4$ where $\kappa$ is given by the dilaton condensate $\la\chi\ra$.  This result comes essentially  from the fact that as pointed out by Yamawaki~\cite{yamawaki}, there is hidden scale symmetry in the strong interactions that also involve hidden local symmetry.  How this comes out in the present paper hangs crucially on what comes out of skyrmion descriptions of dense baryonic matter with the vector mesons and the dilaton present as relevant degrees of freedom.

The behavior of the sound velocity $v_s^2/c^2$ for pure neutron matter predicted by our approach in Fig.~\ref{sound} has a peculiar feature in the vicinity of $n\sim 2n_0$ where the topology change takes place. It starts with $v_s^2/c^2 < 1/3$ at low density $n <2n_0$, goes up to $v_s^2/c^2 > 1/3$ at $n\sim 2n_0$, drops below $1/3$ and then climbs to and asymptotes at 1/3 at higher density $n\gsim 3n_0$.  This feature closely resembles the scenario arrived at by Bedaque and Steiner~\cite{bedaque} in the study of the sound velocity based on their analysis of neutron stars with mass around two solar masses with various phenomenological equations of state. In our theory, there is a rather abrupt changeover of the $bs$HLS Lagrangian parameters due to a topology change (such as the cusp in the skyrmion crystal description), so one might imagine that such a behavior could be an artifact of the sharp transition. One of the characteristic features arising from the topology change  is the stiffening of the symmetry energy at higher density $n > 2n_0$.  It is responsible for the relatively high  proton fraction of nuclear matter in beta equilibrium. It might render the direct URCA process to set in precociously and trigger too rapid a star cooling which might be at odds with observation. If it turns out to be serious,  then that would indicate within our formalism that we need to improve on how the vector manifestation property of the $\rho$ meson sets in at high density.  This matter may be related to what's mentioned in footnote 10 regarding the effective IDD of the gauge coupling constant.  On the other hand, the fact that the dense baryonic matter in our description is in the precursor state to an emerging scale invariance and hence manifesting a conformal-type sound velocity as we are proposing is highly suggestive of the intricate mechanism of the Bedaque-Steiner scenario. What is surprising in our description is, however,  that the ``conformal" sound velocity $v_s^2/c^2=1/3$ sets in so ``precociously" in density and for $\theta_\mu^\mu\neq 0$. In \cite{bedaque}, in contrast,  the matter with $v_s^2/c^2 > 1/3$ should prevail up to density $n \sim 5n_0$, the maximum central density of $\sim 2$-solar mass objects because of the strong hadronic interactions intervening in the phenomenological models they relied on. The conformal velocity should of course appear at very high density.

In the presence of vector mesons, the Lagrangian from which skyrmions arise contains an anomalous term which is present in the Lagrangian even for two flavors and that is the homogeneous Wess-Zumino (hWZ) term~\cite{HLS}. This term does not figure directly in the structure of baryonic matter when the baryon fields are explicitly present: The $V_{lowk}$ RG treatment made above is not affected by this term. However when baryons are generated as skyrmions for dense matter, the hWZ term plays a crucial role because it is through this term that the $\omega$ meson enters in nuclear dynamics. Without it the $\omega$ does not figure in the nuclear interactions. That would be disastrous for the stability of nuclear matter and  for the EoS of dense matter.

The hWZ term consists of three terms with three independent parameters~\cite{HLS}. All three terms need to be included for reliable calculations. For simplicity in notation, let us just take one combination of them in the form ${\cal L}_{\rm hWZ}\propto \omega_\mu B^\mu$ where $B^\mu$ is the topological baryon current. This term is of scale dimension 4, so classically it is scale invariant. However the quantum anomaly introduces $\beta^\prime$ for the anomalous dimension of ${\cal G}$ in the form  ${\cal L}_{\rm hWZ}\to (h_{\rm hWZ} +(1-h_{\rm hWZ})(\chi/f_\sigma)^{\beta^\prime}) {\cal L}_{\rm hWZ}$ where $h$ is an unknown parameter~\cite{MR-omega}. If there were no explicit symmetry breaking of scale symmetry, that is $\beta^\prime$, then there would be no dilaton coupling to the matter fields in the Lagrangian. However it turns out that if $\beta^\prime=0$, the skyrmion matter would diverge~\cite{PRV1}, so the skyrmion matter would make no sense. In order to make the skyrmion structure sensible, it was found to be required that $1\lsim\beta^\prime\lsim 3$ ~\cite{PRV2,MR-omega}.  This means that in order for the $\omega$ meson to figure in the skyrmion matter and assure that $\kappa$ be density-independent,  the explicit scale symmetry must intervene in the hWZ term with a given $\beta^\prime$.  This poses a puzzle as to how this hWZ-term effect is encoded in the IDDs of {\it EFT}$_{\rm bsHLS}$. It is possible that the $\beta^\prime$ is buried in the parameters of the $bs$HLS Lagrangian, without making explicit breaking of scale symmetry in the Lagrangian. This may be related to that if one sets $\sigma=0$ in scale-chiral perturbation theory ($\chi$PT$_\sigma$), one would effectively get the results of three-flavor chiral perturbation theory, $\chi$PT$_3$, which work very well for processes that do not involve scalar channels, with the effects of the trace anomaly hidden in the parameters.

\subsection*{Acknowledgments}
We thank Kyungmin Kim for the TOV code used in the star calculation.
The work of WGP is supported by the Rare Isotope Science Project of Institute for Basic Science funded by Ministry of Science, ICT and Future Planning and National Research Foundation of Korea (2013M7A1A1075764), that of TTSK in part by U.S. Department of Energy  under grant DF-FG02-88ER40388. YLM is supported in part by National Science Foundation of China (NSFC) under Grant No. 11475071, 11547308 and the Seeds Funding of Jilin University. Part of this paper was written while two of the authors (HKL and MR) were visiting RAON/IBS for which the hospitality of Youngman Kim is acknowledged.

\newpage

\appendix

\centerline{\large\bf  APPENDIX}

\setcounter{section}{0}
\renewcommand{\thesection}{\Alph{section}}
\setcounter{equation}{0}
\renewcommand{\theequation}{A.\arabic{equation}}

\begin{center}
\subsection*{The Mean-Field Structure of the Half-Skyrmion Phase}
\end{center}

{In this Appendix, we show that the half-skyrmion phase in the skyrmion-crystal simulation of dense baryonic matter is in a state that can be described entirely by mean fields,  largely undistorted by strong interactions. This resembles Landau-Fermi liquid fixed point theory where the $\beta$ function for the quasiparticle interactions is suppressed. This striking feature was first found in the Skyrme model  with the Atiyah-Manton ansatz in \cite{atiyah-manton-skyrmion}. Here we will show the phenomenon using  the HLS Lagrangian\cite{Ma:2013ooa}\footnote{Since what matters as in the structure of the tensor forces is the topology and symmetry involved, largely independent of strong interactions mediated by non-topological fields, the same argument should apply to the dialton in the half-skrymion phase  in  $s$HLS.}.

We write the chiral field $U$ as $U(\vec{x}) = \phi_0(x,\, y,\, z ) + i \phi^j_\pi(x,\, y,\, z )\tau^j$  with the Pauli matrix $\tau^j$ and $j=1,2,3$. Including $\rho$ and $\omega$, we write the fields placed in the lattice size $L$ as $\phi_{\eta,\, L}(\vec{x}\,)$ with $\eta =0,\, \pi,\, \rho,\, \omega$ and normalize them with respect to their maximum values denoted $\phi_{\eta,L,{\rm max}}$ for given $L$.  It can be shown, as in \cite{atiyah-manton-skyrmion}, with HLS that in the half-skyrmion phase\footnote{The precise value of the half-skyrmion density which depends on the parameters is not important for our discussions.} with $L\lsim L_{1/2}$ where  $L_{1/2}\simeq2.9$ fm, the field configurations are invariant under scaling in density as the lattice is scaled from $L_1$ to $L_2$
\be
\frac{\phi_{\eta,\,L_1}(L_1\vec{t}\,)}{\phi_{\eta,\,L_1,\,{\rm max}}}= \frac{\phi_{\eta,\,L_2}(L_2\vec{t}\,)}{\phi_{\eta,\,L_2,\,{\rm max}}}. \label{scaling}
\ee
Since other fields are quite similar, we only show in  Fig.~\ref{scale_inv} the case of $\phi_{0,\pi}$ for  $\phi_{0,\pi} (t, 0,0)$ vs. $t$ with $t\equiv x/L$.  What is seen there is that denisty-scale invariance sets in for $L\lsim L_{1/2}$. One can see that the field is independent of density in the half-skyrmion phase with $L\lsim L_{1/2}$  whereas for the skyrmion phase with lower density with $L >  L_{1/2}$, it is appreciably dependent on density.
\begin{figure}[h]
\begin{center}
\includegraphics[width=7.0cm]{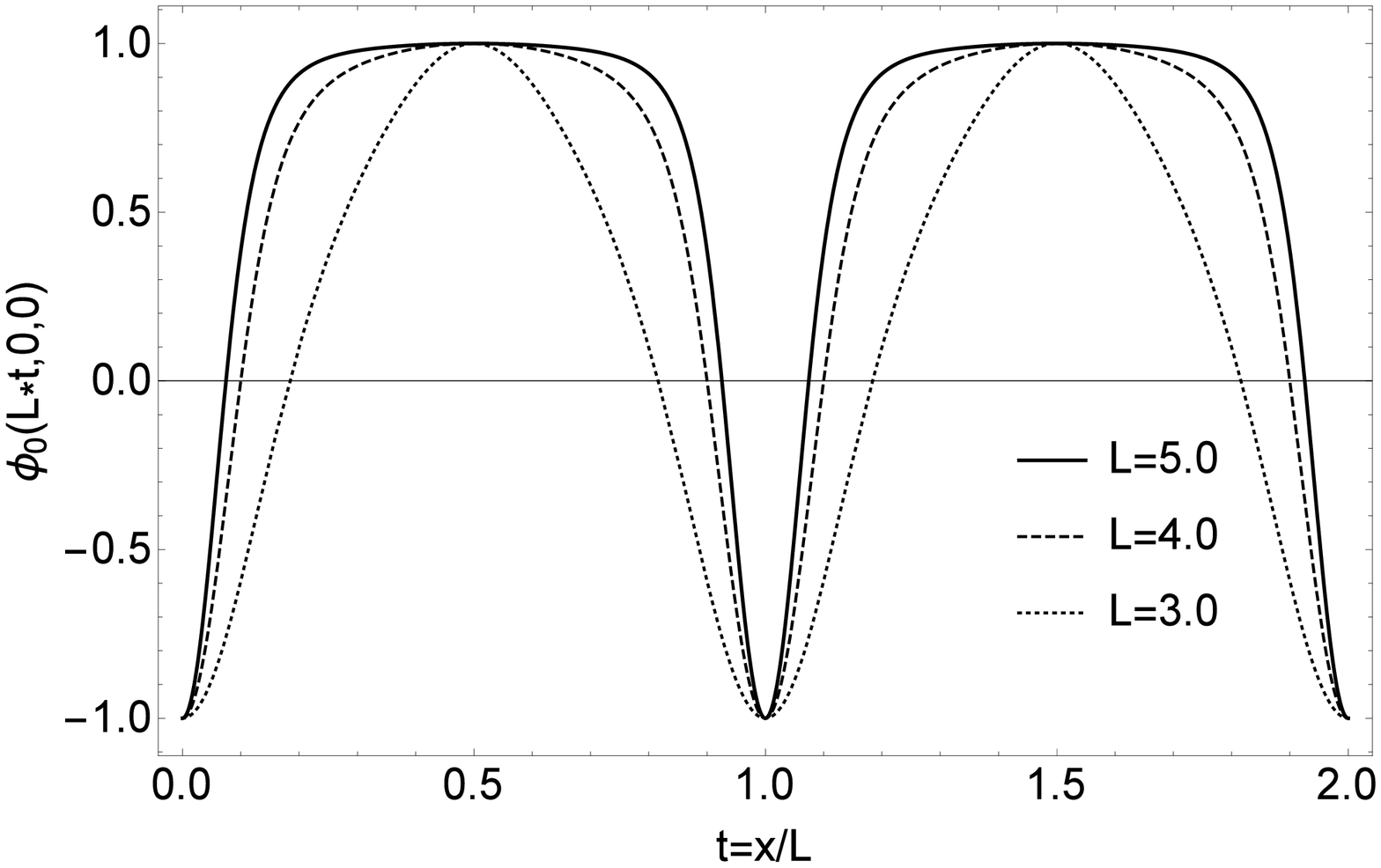}\includegraphics[width=7.0cm]{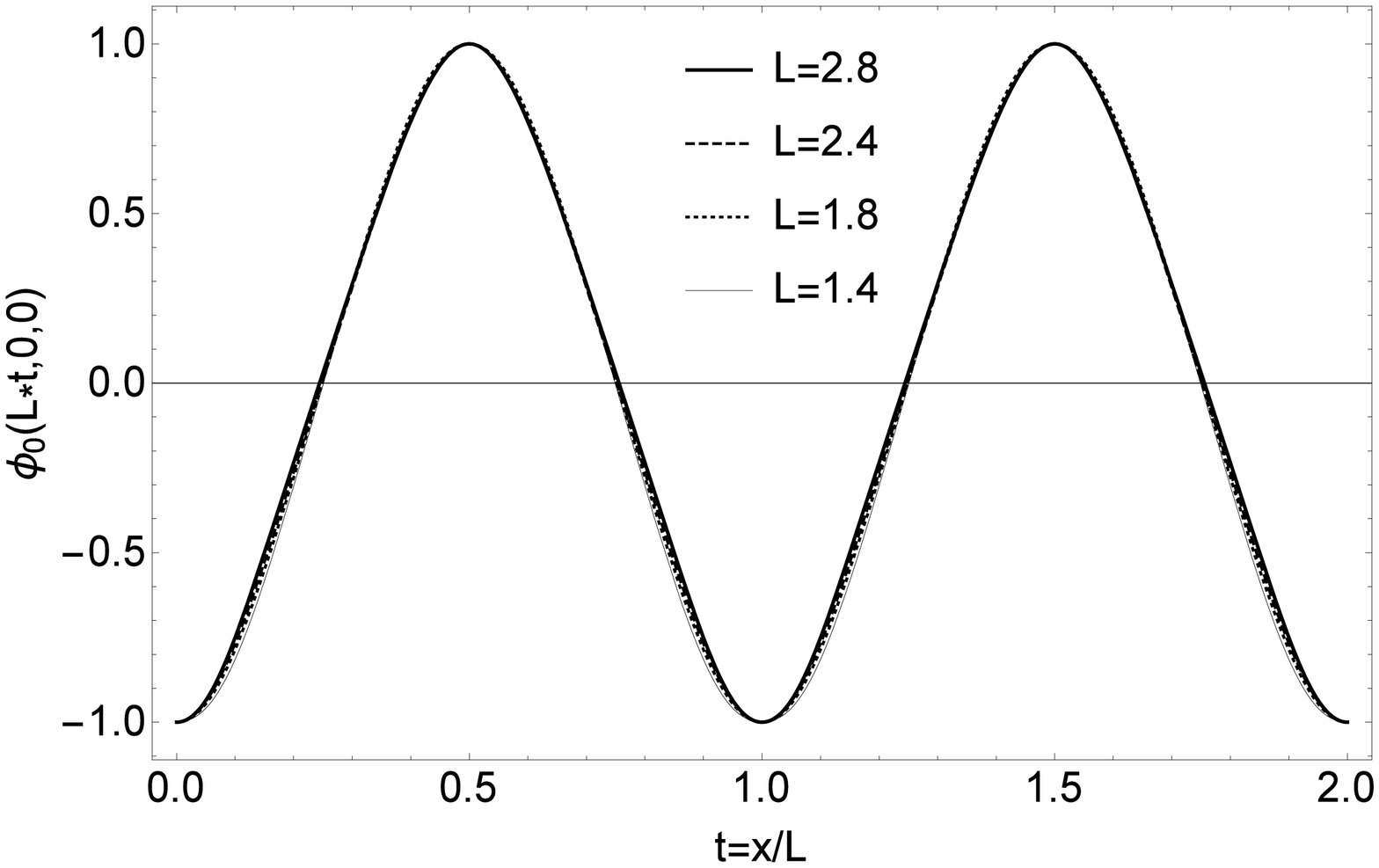}
\includegraphics[width=7.0cm]{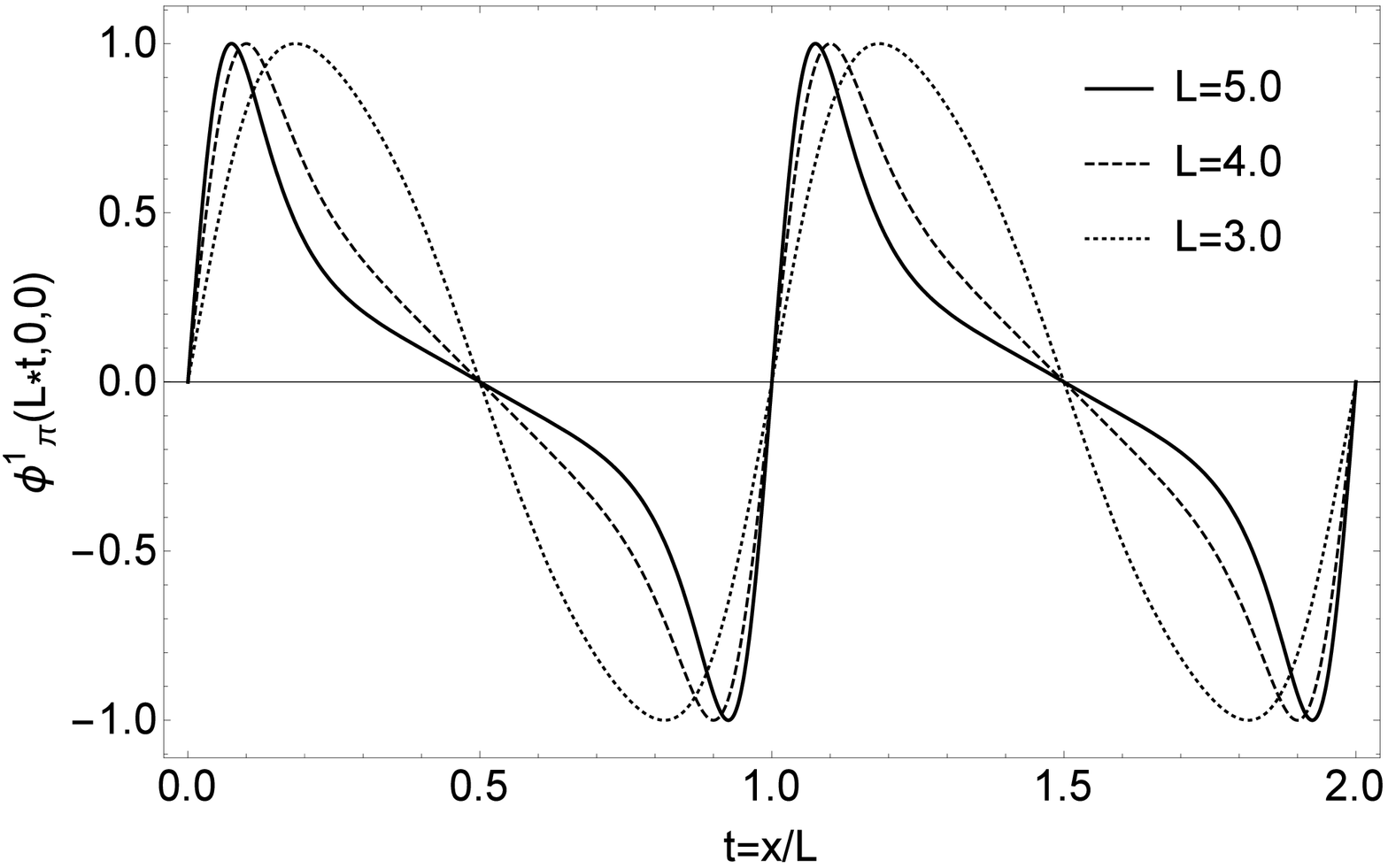}\includegraphics[width=7.0cm]{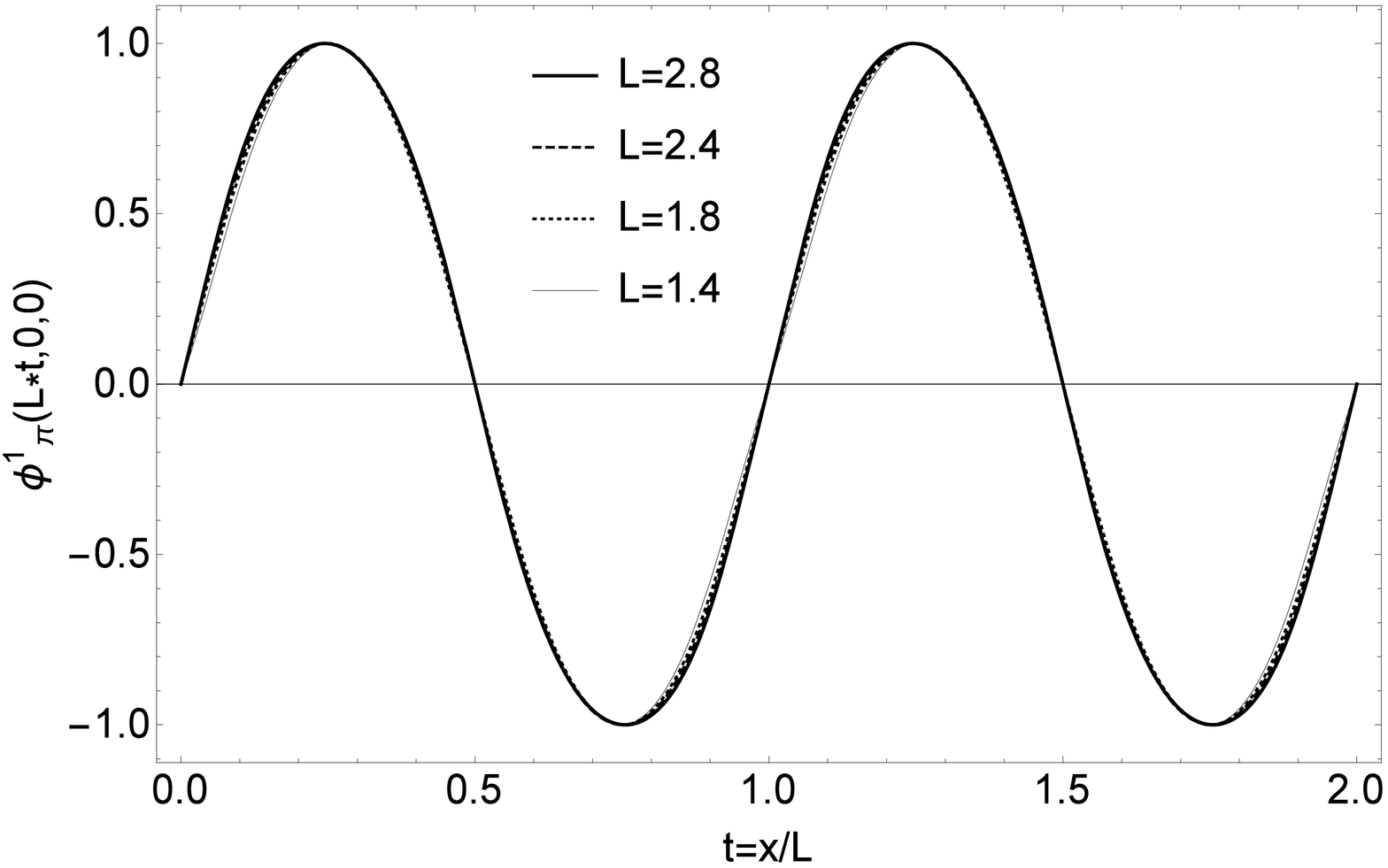}
\caption{  The field configurations $\phi_0$ and $\phi^1_\pi$  as a function of $t = x/L$ along the y = z = 0 line. The maximum values for $\eta=0,\pi$ are $\phi_{0,\,L,\, {\rm max}} = \phi_{\pi,\,L,\, {\rm max}} = 1$. The half-skyrmion phase sets in when $L=L_{1/2} \lsim 2.9\,{\rm fm}$.
 }\label{scale_inv}
 \end{center}
\end{figure}

What does this imply for the energy density?

The energy density for the skyrmion matter put on the lattice of lattice size $L$ can be written as
\begin{equation}
\epsilon = E/A/V(=L^3) = \frac{1}{L^3} \int^L_0 d^3x \sum_{n,\,m} c_{n,m}\, f_{n,m}\left( \vec{\nabla}_x,\, \phi_{\eta,\,L}(\vec{x}\,)\right)\,,
\end{equation}
where $c_{n,m}$ is the coefficient of $f_{n,m}$ which is the function of $\vec{\nabla}_x$ and $\phi_{\eta,\,L}(\vec{x}\,)$ having $n$th power of $\vec{\nabla}_x$ and $m$th power $\phi_{\eta,\,L}(\vec{x}\,)$ with $\nabla_{x,\,j} = \frac{\partial}{\partial\,x^j}$. One can reduce it to
\begin{eqnarray}
\epsilon
&=& \sum_{n,\,m} \left(\frac{1}{L} \right)^n \left(\phi_{\eta,\,L,\, {\rm max}} \right)^m \int^L_0 \frac{d^3x}{L^3}\, c_{n,m}\, f_{n,m}\left( L \vec{\nabla}_x,\, \frac{\phi_{\eta,\,L}(\vec{x}\,)}{\phi_{\eta,\,L,\, {\rm max}}}\right) \\
&=& \sum_{n,\,m} \left(\frac{1}{L} \right)^n \left(\phi_{\eta,\,L,\, {\rm max}} \right)^m \int^1_0 d^3t\, c_{n,m}\, f_{n,m}\left( \vec{\nabla}_t,\, \frac{\phi_{\eta,\,L}(L\vec{t}\,)}{\phi_{\eta,\,L,\, {\rm max}}}\right) \\
&=& \sum_{n,\,m} \left(\frac{1}{L} \right)^n \left(\phi_{\eta,\,L,\, {\rm max}} \right)^m  A_{n,\,m}\,, \label{energy_den}
\end{eqnarray}
 where $A_{n,\,m}$ is a constant independent of the lattice size $L$.

 Calculating the energy density (\ref{energy_den}) in skyrmion-crystal simulations involves field configurations satisfying their equations of motion. Hence (\ref{energy_den}) is a mean field expression. It captures all essential dynamics in terms of the mean fields of each degrees of freedom involved, with residual interactions suppressed. The density dependence lies, apart from the $(1/L)^n$ factor, in the maximum field configuration $\left(\phi_{\eta,\,L,\, {\rm max}} \right)^m$. This implies that in the half-skyrmion phase,  considered to set in at high density, the mean-field structure dominates. This agrees with the lore that at high density -- and in the large $N_c$ limit, the skyrmion crystal picture becomes valid  in QCD. In clear contrast, however, as one can see in Fig.~\ref{scale_inv}, the mean-filed structure breaks down in the lower-density phase with $L > L_{1/2}$. Taking the RMF approximation to be more or less equivalent to Fermi-liquid fixed point theory at high density, one can take the breakdown of the mean-field structure  as a signal for non-Fermi liquid structure. This also agrees with the understanding that the property of low-density baryonic matter -- including nuclear matter -- may be poorly captured in crystal.


\begin{thebibliography}{99}
\bibitem{PKLR}  W.~G.~Paeng, T.~T.~S.~Kuo, H.~K.~Lee and M.~Rho,
  ``Scale-innvariant hidden local symmetry, topology change and dense baryonic matter,''
  Phys.\ Rev.\ C {\bf 93}, no. 5, 055203 (2016).

\bibitem{PR-sound} W.~G.~Paeng and M.~Rho,
  ``Scale-chiral symmetry and the sound velocity in compact-star matter,''
  arXiv:1611.09975 [nucl-th].


\bibitem{Paeng:2013xya}
  W.~G.~Paeng, H.~K.~Lee, M.~Rho and C.~Sasaki,
  ``Interplay between $\omega$-nucleon interaction and nucleon mass in dense baryonic matter,''
  Phys.\ Rev.\ D {\bf 88}, 105019 (2013).

\bibitem{LMR-2016}
  Y.~L.~Li, Y.~L.~Ma and M.~Rho,
  ``Chiral-scale effective theory including a dilatonic meson,''
  arXiv:1609.07014 [hep-ph].

\bibitem{CT}  R.~J.~Crewther and L.~C.~Tunstall,
  ``$\Delta I=1/2$ rule for kaon decays derived from QCD infrared fixed point,''
  Phys.\ Rev.\ D {\bf 91}, no. 3, 034016 (2015).


\bibitem{GS}  M.~Golterman and Y.~Shamir,
  ``Low-energy effective action for pions and a dilatonic meson,''
  Phys.\ Rev.\ D {\bf 94}, no. 5, 054502 (2016).


\bibitem{MR-omega}
  Y.~L.~Ma and M.~Rho,
  ``Scale-chiral symmetry, $\omega$ meson and dense baryonic matter,''
  arXiv:1612.04079 [nucl-th].


\bibitem{yamawaki}  K.~Yamawaki,
  ``Hidden Local Symmetry and Beyond,''
  Int.\ J.\ Mod.\ Phys.\ E {\bf 26}, no. 01n02, 1740032 (2017)




\bibitem{HLS}  M.~Bando, T.~Kugo and K.~Yamawaki,
  ``Nonlinear realization and hidden local symmetries,''
  Phys.\ Rept.\  {\bf 164}, 217 (1988).

\bibitem{HY:PR}  M.~Harada and K.~Yamawaki,
  ``Hidden local symmetry at loop: A New perspective of composite gauge boson and chiral phase transition,''
  Phys.\ Rept.\  {\bf 381}, 1 (2003).


\bibitem{Review-china} Y.~L.~Ma and M.~Rho,
  ``Recent progress on dense nuclear matter in skyrmion approaches,''
  Sci.\ China Phys.\ Mech.\ Astron.\  {\bf 60}, no. 3, 032001 (2017)

\bibitem{haensel}   M.~Fortin, C.~Providencia, A.~R.~Raduta, F.~Gulminelli, J.~L.~Zdunik, P.~Haensel and M.~Bejger,
  ``Neutron star radii and crusts: uncertainties and unified equations of state,''
  Phys.\ Rev.\ C {\bf 94}, no. 3, 035804 (2016).



\bibitem{paritydoublet} C.~E.~Detar and T.~Kunihiro,
  ``Linear $\sigma$ model with parity doubling,''
  Phys.\ Rev.\ D {\bf 39}, 2805 (1989).

\bibitem{Harada:2015lma}
  M.~Harada, H.~K.~Lee, Y.~L.~Ma and M.~Rho,
  ``Inhomogeneous quark condensate in compressed Skyrmion matter,''
  Phys.\ Rev.\ D {\bf 91}, no. 9, 096011 (2015)  doi:10.1103/PhysRevD.91.096011  [arXiv:1502.02508 [hep-ph]].  

\bibitem{ma-harada2016} M.~Harada, Y.~L.~Ma, D.~Suenaga and Y.~Takeda,
  ``Effect of omega meson on the heavy-light mesons with chiral partner structure in dense matter,''
  arXiv:1612.03496 [hep-ph].


\bibitem{harland}  D.~Harland,
  ``The Skyrme model and chiral perturbation theory,''
  arXiv:1612.07259 [hep-th].


\bibitem{matsui} T.~Matsui,
  ``Fermi liquid properties of nuclear matter in a relativistic mean-field Theory,''
  Nucl.\ Phys.\ A {\bf 370}, 365 (1981).

\bibitem{chaejun} C.~Song, D.~P.~Min and M.~Rho,
  ``Thermodynamic properties of effective chiral Lagrangians with Brown-Rho scaling,''
  Phys.\ Lett.\ B {\bf 424}, 226 (1998)

\bibitem{holt-kaiser} J.~W.~Holt and N.~Kaiser,
  ``Equation of state of nuclear and neutron matter at third-order in perturbation theory from chiral EFT,''
  arXiv:1612.04309 [nucl-th].



\bibitem{holt-weise-c14}  J.~W.~Holt, N.~Kaiser and W.~Weise,
  ``Density-dependent nuclear interactions and the beta decay of 14C: chiral three-nucleon forces and Brown-Rho scaling,''
    arXiv:1011.6623 [nucl-th].

\bibitem{dongetal} H.~Dong, T.~T.~S.~Kuo, H.~K.~Lee, R.~Machleidt and M.~Rho,
  ``Half-skyrmions and the equation of state for compact-star matter,''
  Phys.\ Rev.\ C {\bf 87}, 054332 (2013) .

\bibitem{BR:DD}  G.~E.~Brown and M.~Rho,
  ``Double decimation and sliding vacua in the nuclear many body system,''
  Phys.\ Rept.\  {\bf 396}, 1 (2004).


\bibitem{vlowk-fermiliquid} J.~W.~Holt, G.~E.~Brown, J.~D.~Holt and T.~T.~S.~Kuo,
  ``Nuclear matter with Brown-Rho-scaled Fermi liquid interactions,''
  Nucl.\ Phys.\ A {\bf 785}, 322 (2007).

\bibitem{cusp} H.~K.~Lee, B.~Y.~Park and M.~Rho,
  ``Half-skyrmions, tensor forces and symmetry energy in cold dense matter,''
  Phys.\ Rev.\ C {\bf 83}, 025206 (2011)
  Erratum: [Phys.\ Rev.\ C {\bf 84}, 059902 (2011)].


\bibitem{Song:2000cu}
  C.~Song,
  ``Dense nuclear matter: Landau Fermi liquid theory and chiral Lagrangian with scaling,''
  Phys.\ Rept.\  {\bf 347}, 289 (2001).

 \bibitem{shankar}
  R.~Shankar,
 ``Renormalization group approach to interacting fermions,''
  Rev.\ Mod.\ Phys.\  {\bf 66}, 129 (1994).


%
\bibitem{yamazaki}
  P.~Kienle and T.~Yamazaki,
  ``Pions in nuclei, a probe of chiral symmetry restoration,''
  Prog.\ Part.\ Nucl.\ Phys.\  {\bf 52}, 85 (2004).

\bibitem{FRG-Weise}  M.~Drews and W.~Weise,
 ``Functional renormalization group studies of nuclear and neutron matter,''
 Prog. Part. Nucl. Phys. {\bf 93},  69 (2017).

\bibitem{Danielewicz:2002pu}
  P.~Danielewicz, R.~Lacey and W.~G.~Lynch,
  ``Determination of the equation of state of dense matter,''
  Science {\bf 298}, 1592 (2002).

%
\bibitem{Li:2005jy}
  B.~A.~Li and L.~W.~Chen,
  ``Nucleon-nucleon cross sections in neutron-rich matter and isospin transport in heavy-ion reactions at intermediate energies,''
  Phys.\ Rev.\ C {\bf 72}, 064611 (2005)
  doi:10.1103/PhysRevC.72.064611
  [nucl-th/0508024].

  %
\bibitem{Tsang:2008fd}
  M.~B.~Tsang et al.,
  ``Constraints on the density dependence of the symmetry energy,''
  Phys.\ Rev.\ Lett.\  {\bf 102}, 122701 (2009)
  [Int.\ J.\ Mod.\ Phys.\ E {\bf 19}, 1631 (2010)].


 \bibitem{Antoniadis:2013pzd}
  J.~Antoniadis {\it et al.},
  ``A massive pulsar in a compact relativistic binary,''
  Science {\bf 340}, 6131 (2013).



\bibitem{DH} F.~Douchin and P.~Haensel,
  ``A unified equation of state of dense matter and neutron star structure,''
  Astron.\ Astrophys.\  {\bf 380}, 151 (2001)

\bibitem{AP3}A.~Akmal, V.~R.~Pandharipande and D.~G.~Ravenhall,
  ``The Equation of state of nucleon matter and neutron star structure,''
  Phys.\ Rev.\ C {\bf 58}, 1804 (1998)

\bibitem{MPA}H.~Müther, M.~Prakash and T.~L.~Ainsworth,
  ``The nuclear symmetry energy in relativistic Brueckner-Hartree-Fock calculations,''
  Phys.\ Lett.\ B {\bf 199}, 469 (1987).

\bibitem{PRD/81/123016} T.~Hinderer, B.~D.~Lackey, R.~N.~Lang and J.~S.~Read,
  ``Tidal deformability of neutron stars with realistic equations of state and their gravitational wave signatures in binary inspiral,''
  Phys.\ Rev.\ D {\bf 81}, 123016 (2010)
  doi:10.1103/PhysRevD.81.123016


\bibitem{read} J.~S.~Read {\it et al.},
  ``Matter effects on binary neutron star waveforms,''
  Phys.\ Rev.\ D {\bf 88}, 044042 (2013)

\bibitem{HKSS} K.~Hotokezaka, K.~Kyutoku, Y.~i.~Sekiguchi and M.~Shibata,
  ``Measurability of the tidal deformability by gravitational waves from coalescing binary neutron stars,''
  Phys.\ Rev.\ D {\bf 93}, no. 6, 064082 (2016)

\bibitem{KLL}  K.~Kim, H.~K.~Lee and J.~Lee,
  ``Compact Star Matter: EoS with New Scaling Law,''
  Int.\ J.\ Mod.\ Phys.\ E {\bf 26}, no. 01n02, 1740011 (2017)



\bibitem{bedaque} P.~Bedaque and A.~W.~Steiner,
  ``Sound velocity bound and neutron stars,''
  Phys.\ Rev.\ Lett.\  {\bf 114},  031103 (2015).

\bibitem{radice} D. Radice et al., arXive:1612.06429


\bibitem{Lee:2003eg}
  H.~J.~Lee, B.~Y.~Park, M.~Rho and V.~Vento,
  ``Sliding vacua in dense skyrmion matter,''
  Nucl.\ Phys.\ A {\bf 726}, 69 (2003).

\bibitem{PRV1} B.~Y.~Park, M.~Rho and V.~Vento,
  ``Vector mesons and dense skyrmion matter,''
  Nucl.\ Phys.\ A {\bf 736}, 129 (2004).




  \bibitem{weinberg}
  S.~Weinberg,
  ``Mended symmetries,''
  Phys.\ Rev.\ Lett.\  {\bf 65}, 1177 (1990);
  `Unbreaking symmetries,''
  Conf.\ Proc.\ C {\bf 930308}, 3 (1993).
\bibitem{fukushima}
 K.~Fukushima and T.~Kojo,
  ``The quarkyonic star,''
  Astrophys.\ J.\  {\bf 817}, no. 2, 180 (2016).


\bibitem{PRV2}  B.~Y.~Park, M.~Rho and V.~Vento,
  ``The role of the dilaton in dense skyrmion matter,''
  Nucl.\ Phys.\ A {\bf 807}, 28 (2008)


\bibitem{atiyah-manton-skyrmion}  B.~Y.~Park, D.~P.~Min, M.~Rho and V.~Vento,
  ``Atiyah-Manton approach to skyrmion matter,''
  Nucl.\ Phys.\ A {\bf 707}, 381 (2002).

\bibitem{Ma:2013ooa}
  Y.~L.~Ma, M.~Harada, H.~K.~Lee, Y.~Oh, B.~Y.~Park and M.~Rho,
  ``Dense baryonic matter in the hidden local symmetry approach: Half-skyrmions and nucleon mass,''
  Phys.\ Rev.\ D {\bf 88}, no. 1, 014016 (2013)
  Erratum: [Phys.\ Rev.\ D {\bf 88}, no. 7, 079904 (2013)].



\end{thebibliography}
\end{document}